\documentclass[fleqn,usenatbib]{mnras}

\usepackage{newtxtext,newtxmath}

\usepackage[T1]{fontenc}
\usepackage{ae,aecompl}

\usepackage{graphicx}	
\usepackage{amsmath}	
\usepackage{amssymb}	

\usepackage[usenames,dvipsnames,svgnames,hyperref]{xcolor}

\usepackage{comment}

\newcommand{\msuni}{\mathrm{M}_\odot}

\newcommand{\mstot}{M_{*,\mathrm{tot}}}
\newcommand{\nsats}{N_\mathrm{sats}}
\newcommand{\fmihsc}{f_{M_{*,\mathrm{IHSC}}}}
\newcommand{\mscnd}{M_{*,\mathrm{scnd}}}
\newcommand{\mctrl}{M_{*,\mathrm{ctrl}}}
\newcommand{\mihsc}{M_{*,\mathrm{IHSC}}}
\newcommand{\msats}{M_{*,\mathrm{sats}}}
\newcommand{\mfsm}{f_{M_{*,\mathrm{IHSC}}} - M_{*}}
\newcommand{\mvir}{M_{200c}}
\newcommand{\rvir}{R_{200c}}

\title[Intra-Halo Stellar Component]{From Stellar Halos to Intracluster Light:
the physics of the Intra-Halo Stellar Component in cosmological hydrodynamical
simulations}

\author[R. Ca\~nas et al.]{Rodrigo 
Ca\~nas,$^{1,2}$\thanks{E-mail: rodrigo.canas@icrar.org (RC)},
Claudia del P. Lagos$^{1,2}$,
Pascal J. Elahi$^{1,2}$,
Chris Power$^{1,2}$,
\newauthor
Charlotte Welker$^{1,2,3}$,
Yohan Dubois$^{4,5,6}$,
and Christophe Pichon$^{4,5,6}$
\\
$^{1}$International Centre for Radio Astronomy Research, University of Western
Australia, 35 Stirling Highway, Crawley, WA 6009,\\
Australia\\
$^{2}$ARC Centre of Excellence for All Sky Astrophysics in 3 Dimensions (ASTRO 3D)\\
$^{3}$Department of Physics and Astronomy, McMaster University, Hamilton, Ontario, Canada\\
$^{4}$CNRS and UPMC Univ. Paris 06, UMR 7095, Institut d'Astrophysique de Paris, 98 bis Boulevard Arago, F-75014 Paris, France\\
$^{5}$Institute for Astronomy, University of Edinburgh, Royal Observatory, Blackford Hill, Edinburgh, EH9 3HJ, United Kingdom\\
$^{6}$Korea Institute of Advanced Studies (KIAS) 85 Hoegiro, Dongdaemun-gu, Seoul, 02455, Republic of Korea
}

\date{Accepted XXX. Received YYY; in original form ZZZ}

\pubyear{2019}

\begin{document}
\label{firstpage}
\pagerange{\pageref{firstpage}--\pageref{lastpage}}
\maketitle

  \begin{abstract}
    We study the Intra-Halo Stellar Component (IHSC) of Milky
    Way-mass systems up to galaxy clusters in the Horizon-AGN
    cosmological hydrodynamical simulation.
    We identify the IHSC using an improved phase-space galaxy
    finder algorithm which provides an adaptive,  physically
    motivated and shape-independent definition of this stellar
    component, {that can be applied to halos of arbitrary masses}.
    We explore the IHSC mass fraction - total halo's stellar
    mass, $\mfsm$, relation and the physical drivers of its
    scatter.
    We find that on average the $\fmihsc$ increases with total 
    stellar mass, with the scatter decreasing strongly with mass
    from $2$~dex at $\mstot \simeq 10^{11}\,\msuni$ to $0.3$~dex
    at group masses. 
    At high masses, $\mstot > 10^{11.5}\,\msuni$,
    $\fmihsc$ increases with the number of substructures, and with
    the mass ratio between the central galaxy and largest satellite,
    at fixed $\mstot$.
    From mid-size groups and systems below $\mstot < 10^{12}\,\msuni$,
    we find that the central galaxy's stellar rotation-to-dispersion
    velocity ratio, $V/\sigma$, displays the strongest 
    (anti)-correlation with $\fmihsc$ at fixed $\mstot$ of all the
    galaxy and halo properties explored, transitioning from
    $\fmihsc< 0.1$\% for high $V/\sigma$, to $\fmihsc\approx 5$\% for
    low $V/\sigma$ galaxies. 
    By studying the $\fmihsc$ temporal evolution, we find that, in
    the former, mergers not always take place, but if they did, they
    happened early ($z > 1$), while the high $\fmihsc$
    population displays a much more active
    merger history.
    In the case of massive groups and galaxy clusters,
    $\mstot \gtrsim 10^{12}\,\msuni$, a fraction $\fmihsc\approx 10-20$\%
    is reached at $z\approx 1$  and then they evolve across lines of
    constant $\fmihsc$ modulo some small perturbations.
    Because of the limited simulation's volume, the
    latter is only tentative and requires a larger sample
    of simulated galaxy clusters to confirm.
  \end{abstract}




  \section{Introduction}
  \label{sec:intro}
    In the hierarchical formation scenario, large galaxies 
    are assembled via a sequence of interactions and mergers
    with smaller galaxies \citep{White1978}.
    During such events, tidal forces strip stars from these smaller,
    satellite galaxies, which become part of the more massive, 
    central galaxy, or are deposited
    in its outskirts in the form of streams, shells, and a
    diffuse component \citep[e.g.][]{Zwicky1952,Toomre1972,Barnes1991,Mihos1996}.
    The properties of these stellar remnants should contain
    important information about the assembly history and dynamical
    age of Milky Way-like systems
    \citep[e.g.][]{Ibata2005,MartinezDelgado2008,McConnachie2009,Watkins2015,
    Merritt2016,Monachesi2016a},
    galaxy groups \citep[e.g.][]{DaRocha2005,Durbala2008},
    and galaxy clusters \citep[e.g.][]{Mihos2005,Montes2014,Mihos2017,
    Morishita2017}.
    However, the study of such stellar remnants is a complicated task
    because of their diffuse nature, which leads them to have low surface
    brightnesses. This has led to the development of
    multiple techniques to study them in both observations and
    simulations.

    \medskip
    
    From the observational perspective, this requires the use of telescopes and
    techniques to detect low surface-brightness
    (LSB) features, as well as careful treatment of noise, scattered
    light, masking of foreground and background objects, and instrument
    systematics that can affect measurements in this faint regime
    \citep[see e.g.][]{Ibata2005,Mihos2005,McConnachie2009,Barker2009,MartinezDelgado2010,Abraham2014,Monachesi2016a}.
    Moreover, the definition of the diffuse stellar component is likely 
    to depend on characteristics of the telescope and observation,
    as well as the system of interest.
    For studies in which individual stars can be resolved, the diffuse
    stellar outskirts are often defined by their location.
    Specifically, studies of Milky Way-like galaxies measure
    stellar halos using Asymptotic Giant Branch (AGB) stars along
    the minor axis of the central galaxy
    \citep[e.g.][]{Mouhcine2005,Greggio2014,Monachesi2016a}, and along
    the major axis at sufficiently large distances from the galaxy
    \citep{Monachesi2016a}.
    In contrast, studies that use integrated light either treat galaxies
    on an object by object basis \citep[e.g.][]{Mihos2005,Krick2007,MartinezDelgado2010,Merritt2016}, 
    or stack samples of galaxies \citep[e.g.][]{Zibetti2005,DSouza2014,Zhang2018}.
    In the first approach, the light from a galaxy is separated from
    that of the diffuse component by fitting single or multiple S\'ersic
    profiles \citep{Sersic1963}, and the diffuse envelope is treated as either
    the outermost component \citep[e.g.][]{DSouza2014,Zhang2018} or is
    defined by the excess light or mass from fits to the inner regions
    of the galaxies \citep[e.g.][]{Merritt2016,Morishita2017}.
    In the second, a surface brightness threshold separates the light 
    from galaxies and from the diffuse envelope
    \citep{Feldmeier2004,Zibetti2005,Montes2014,Burke2015}, which we note is 
    a method also adopted by some numerical studies \citep[e.g.][]{Rudick2006}.
    \medskip

    From the theoretical perspective, reliable predictions of stellar halos
    and Intracluster Light (ICL) require simulations to
    realistically trace the accretion and subsequent disruption of 
    the satellite galaxies that give rise to these structures, and
    to resolve the sparsely populated outskirts of galaxies.
    Early studies of stellar halos, such as \citet{Bullock2005}
    and \citet{Gauthier2006}, used idealised non-cosmological $N$-body 
    simulations with satellite populations whose properties (e.g.
    orbital parameters and accretion histories) informed by cosmological 
    simulations. Others, e.g. \citet{Rudick2006}, used dark matter halos drawn 
    from cosmological $N$-body simulations, populated with galaxies using an 
    occupation distribution formalism, to study the formation and evolution 
    of the ICL.
    \citet{Cooper2010} studied stellar halo formation in a self-consistent
    cosmological context by coupling a semi-analytical model to a cosmological 
    $N$-body simulation, in which subhalos' most bound dark 
    matter particles are tagged and used as dynamical tracers of the stellar 
    populations predicted, to predict the properties of stellar halos; here it
    was assumed that stellar halos are composed of stellar particles that were 
    accreted from satellites and reside outside a spherical aperture of 3 kpc
    \citep[see also][]{Cooper2013,Cooper2015}.
        
    The current generation of cosmological hydrodynamical simulations 
    \citep[e.g.][]{Dubois2014,Vogelsberger2014,Schaye2015} are now
    sufficient to model self-consistently the formation and hierarchical 
    assembly of statistical samples of galaxies, and consequently 
    track the formation of stellar streams, shells, and halos. This provides
    important new insights into formation processes \citep[e.g. in-situ vs ex-situ halos; cf.][]{Font2011}, but it also complicates the separation of galaxies 
    from their stellar components.
    In uniform resolution cosmological boxes, this has led 
    the diffuse component to be defined as the stellar material
    outside spherical apertures, which can be either fixed
    \citep[e.g.][]{Font2011,Pillepich2018b} or dependent upon the mass
    distribution of the system \citep[e.g.][]{Pillepich2014,Elias2018}.
    In zoom cosmological simulations of late-type galaxies, a more ad hoc 
    approach has been used. For example, \citet{Pillepich2015} used a cylindrical
    volume to separate galaxies from their stellar halos, while
    \citet{Monachesi2019} used rectangular windows, mirroring an observational
    2D approach, to separate components.
    While such spatial definitions are simple and easy to compare
    between studies, they ignore the sometimes complex galaxy
    morphology, and, most importantly, they do not exploit the
    velocity information that is available in simulations.

    However, this is not the case for all existing methods in 
    the literature.
    For example, at galaxy clusters scales, velocity information
    has been used to separate the brightest galaxy cluster (BCG)
    from the ICL, either by comparing particles' binding energy in
    which the ICL is the stellar material bound to the cluster but not
    to a particular galaxy
    \citep[e.g.][]{Murante2004,Murante2007,Rudick2011},
    or by using its kinematics and fitting Maxwellian
    distributions to the total velocity distribution being
    the diffuse component the one with the largest dispersion
    \citep[e.g.][]{Puchwein2010,Dolag2010,Cui2014,Remus2017}.
    These are both physically motivated definitions; however, in applying
    the first method, it is not clear how to properly disentangle the
    contribution from individual cluster members to the global
    gravitational potential \citep{Murante2004,Rudick2011}, while in applying
    the second method, how particles are assigned to the distributions 
    is not unique \citep{Dolag2010,Cui2014} and the number of distributions
    needed to describe the system can vary between clusters
    \citep[][]{Remus2017}.
    \medskip
    
    Studies of the diffuse stellar component in the literature have 
    focused on Milky Way-like galaxies, in which case the diffuse stellar
    component is equivalent to a stellar halo, or on galaxy groups
    and clusters, in which case it is the IGL and ICL, respectively.
    This, in addition to the variety of definitions of the diffuse
    stellar component and the techniques used to identify it,
    has limited our understanding of how it is built across the dynamic 
    range of galaxy formation.
    This critical limitation has resulted in a 
    disconnect between the study of the assembly of galaxies and the 
    build up of the diffuse stellar component, despite hierarchical
    growth underpinning both.
    Some theoretical studies have addressed this issue either by 
    applying the same technique or definition to systems of a wide mass
    range
    \citep[e.g.][]{Cooper2010,Cooper2013,Cooper2015,Pillepich2018b}, 
    or using an adaptive definition of what the diffuse component
    is \citep[e.g.][]{Pillepich2014,Elias2018}.
    However, there is still not yet a physically motivated adaptive
    definition that can be reliably applied to the whole dynamic range
    of systems resolved by cosmological hydrodynamical simulations.
    %
    
    In this work we present the first results of a new method to
    identify the diffuse stellar component in simulations based
    on the galaxy finding algorithm described in \citet{Canas2019}
    developed within the phase-space structure finder code
    {\sc velociraptor} \citep{Elahi2011,Elahi2019a}.
    In this method, the diffuse stellar component is defined as 
    kinematically hot stellar particles that are distinct from the
    stellar components of phase-space overdense galaxies.
    The algorithm is adaptive and capable of
    separating the diffuse component in fairly isolated systems
    as well as in complex ones such as galaxy groups and clusters.
    We have decided to refer to the diffuse stellar component as 
    the Intra-Halo Stellar Component (IHSC).
    \medskip
    
    This study aims to characterize the mass content of the
    IHSC across different mass ranges and epochs by exploring to
    detail the IHSC mass fraction - stellar mass relation,
    $\mfsm$.
    In particular, we want to understand the origin
    of the large scatter observed in the IHSC mass fraction
    of Milky Way-mass like galaxies (over $\sim$2 dex difference
    from peak-to-peak), which has been measured recently by 
    \citet{Merritt2016} using integrated light observations of
    the Dragonfly Telephoto Array \citep{Abraham2014}, and later
    confirmed by \citet{Harmsen2017} with stellar counts observations
    of the GHOSTS Survey \citep{RadburnSmith2011}.
    We also want to understand whether the ICL
    fraction\footnote{Referring either to light or mass,
    which can be interchangeable if a constant
    mass-to-light ratio, $M/L$, is assumed.} correlates with
    cluster mass \citep{Murante2007,Rudick2011}
    or not \citep{Krick2007,Contini2014,Cui2014}, as well as, 
    if its evolution is strong \citep{Burke2015}, or rather
    weak or nonexistent \citep{Krick2007,Rudick2011,Montes2018}; 
    this is an unresolved problem, from both observational 
    and theoretical perspectives.
    In this first study we use the Horizon-AGN
    simulation \citep{Dubois2014}, to address this problem.
    This state-of-the art simulation has a volume big enough
    to contain galaxy groups and low-mass galaxy clusters 
    ($\sim 400$ halos with $\mvir > 10^{13}\msuni$), as well 
    as, enough resolution to explore Milky Way-like systems.
    %

    This paper is organised as follows.
    We first describe in Section~\ref{sec:methodology} the Horizon-AGN
    simulation and the algorithm used to identify galaxies and the IHSC,
    as well as a description of how properties used throughout the 
    paper are calculated.
    In Section~\ref{sec:fmihsc} we show visually the IHSC for systems
    ranging in mass from the Milky Way up to galaxy clusters.
    We also describe the $z=0$ $\mfsm$ relation and show how the mass
    content in galaxies and the IHSC is affected by parameters of
    our identification method, as well as how our method compares
    to spherical aperture definitions of the IHSC.
    In Section~\ref{sec:unveiling}, we explore the origin
    of the scatter observed in the $\mfsm$ relation and correlations
    with galaxy properties.
    In Section~\ref{sec:ihscevol} we investigate the evolution of the
    IHSC for individual systems, as well as entire galaxy populations.
    Finally in Section~\ref{sec:conclusions} we present a 
    summary and conclusions of this work.
    In Appendix~\ref{appndx:fixedapertures} we present a comparison
    between our method to separate the IHSC and widely used
    definitions in the literature, and also show the mass fraction
    in the IHSC as a function of halo mass.

  \section{Methodology}
  \label{sec:methodology}
    \subsection{Horizon-AGN Simulation}
    \label{sec:horizonagn}
      Horizon-AGN, first described in \citet{Dubois2014},
      is a state-of-the-art hydrodynamical simulation of
      a statistically representative volume of the universe
      in a periodic box of $L_\mathrm{box} = 100\,h^{-1}$ Mpc 
      on each side, with a $\Lambda$ cold dark matter ($\Lambda$CDM)
      cosmology.
      It adopts values of a total matter density
      $\Omega_\mathrm{m} = 0.272$, dark energy density
      $\Omega_\Lambda = 0.728$, amplitude of the linear power
      spectrum $\sigma_8 = 0.81$, baryon density
      $\Omega_\mathrm{b} = 0.045$, Hubble constant
      $H_0 =70.4$ km\,s$^{-1}$\, Mpc $^{-1}$, and spectral
      index $n_s = 0.967$, in concordance to results from the 
      \emph{Wilkinson Microwave Anisotropy Probe 7} 
      \citep[WMAP7,][]{Komatsu2011}.
      %
      
      The simulation follows the formation and evolution of
      galaxies using the adaptive mesh refinement (AMR)
      code {\sc ramses} \citep{Teyssier2002}, with a 
      total of 1024$^3$ dark matter particles with mass 
      $M_\mathrm{dm} = 8 \times 10^7 \msuni$.
      It has an initial number of 1024$^3$ gas cells, which
      are refined up to seven times reaching a maximum physical
      resolution of $\sim$1 kpc.
      It includes gas cooling, heating from a uniform 
      redshift-dependent UV background, star formation, stellar
      feedback driven by supernovae (SNe) Type Ia and II, and
      stellar winds.
      Black holes (BHs) grow according to a 
      Bondi-Hoyle-Lyttleton accretion scheme capped at the
      Eddington accretion rate, and a two-mode AGN
      feedback is explicitly implemented as
      an isotropic thermal energy injection at
      accretion rates greater than 1\% the Eddington
      accretion, and as a bipolar outflow otherwise
      \citep[see][for further details] 
      {Dubois2010,Dubois2012,Dubois2014};
      these implementations produce BH populations
      that reproduce the observed evolution of the BH mass
      density and mass functions \citep{Volonteri2016}.
      %

      Horizon-AGN produces galaxy 
      populations whose luminosity and stellar mass
      functions, as well as star formation main 
      sequence, are in good agreement with the
      observed ones \citep{Kaviraj2017}.
      It has also been used to demonstrate that AGN
      feedback is crucial to produce the
      observed morphology diversity and kinematic
      properties of massive galaxies \citep{Dubois2016},
      and to show how AGN feedback affects the total
      density profile of galaxies and dark matter halos 
      \citep{Peirani2017,Peirani2019}.
      In addition, the simulation has been used to investigate the
      alignment between cosmic web filaments and the 
      spins of galaxies, and how mergers change the spin 
      orientation of galaxies \citep{Dubois2014,Welker2014}.

  \subsection{Identification of structures}
  \label{sec:galihscid}
    Structure is identified using the code {\sc velociraptor}, first
    introduced in \citet{Elahi2011} and subsequently upgraded in 
    \citet[][]{Canas2019} and \citet{Elahi2019a}.
    Here we describe the identification process for dark
    matter halos, galaxies and the IHSC. 

    \subsubsection{Galaxies}
      Galaxies are identified using a phase-space (6D) 
      Friends-of-Friends \citep[FOF]{Davis1985} search to
      identify phase-space dense structures.
      The algorithm is described in detail in \citet{Canas2019},
      but here we summarise key features.
      The first step consists of obtaining a configuration space
      (3D) FOF, which is done using the commonly adopted $b = 0.2$
      dark matter inter-particle spacing as linking length, $l_x$.
      Subsequently a 6DFOF search is done, where $l_x$ is shrunk
      as galaxies are expected to be more concentrated 
      than dark matter halos;
      the velocity space linking length is chosen to be equal to
      the 3DFOF object velocity dispersion, $l_v = \sigma_v$.
      Substructures are identified by performing an iterative
      6DFOF search, as is used for core-finding in dark matter halos.
      A key feature of this process for galaxies is the selection
      of the initial and iterative parameters used for this
      search, and properly assigning particles to each of these
      cores.
      The most massive galaxy in each 3DFOF object is considered
      to be the central, and the remaining objects to be 
      substructures/satellites.

    \subsubsection{Dark matter halos}
      Dark matter halos and their substructure can also be
      identified with {\sc velociraptor}.
      However, because we focus on the stellar mass
      content in galaxies and the IHSC, dark matter halos will
      only be used to refer to the total matter content
      (stars + BH + gas + dark matter) in the region of interest.
      Halos are therefore defined as spherical regions centred at
      the centre-of-mass of the most massive galaxy of a FOF
      object, in which the total mass average overdensity is
      200 times the critical density of the universe.
      Throughout the paper we use the subscript ${200c}$ to
      refer to properties of these objects, e.g $M_{200c}$.

    \subsubsection{Defining the IHSC}
    \label{sec:ihscdef}

      \begin{figure*}
        \includegraphics[width=0.89\textwidth]{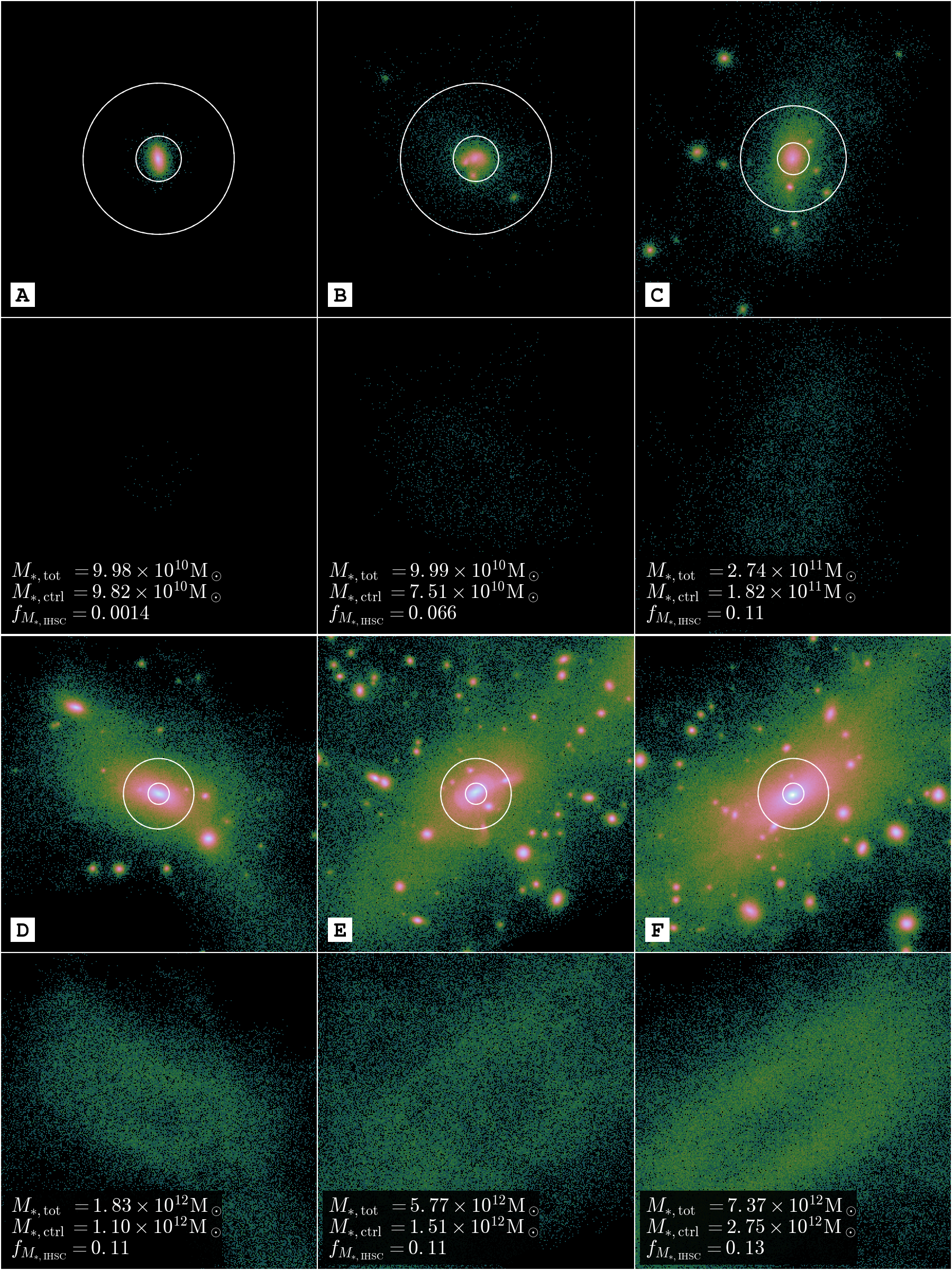}
        \caption{Projected stellar mass density of all the stellar particles
                 inside 3DFOF objects (first and third row), and its
                 IHSC as identified using {\sc velociraptor} (second and 
                 fourth row).
                 3DFOF total stellar mass, $\mstot$, central galaxy
                 stellar mass, $\mctrl$, and IHSC mass fraction 
                 $\fmihsc = \mihsc/\mstot$ are shown for each object.
                 Circles show 30 and 100 kpc apertures commonly
                 used in the literature to separate the central galaxy
                 from the stellar halo/ICL.
                 }
        \label{fig:ihscvis}
      \end{figure*}

      The IHSC is composed
      of all the background particles that were not assigned to
      galaxies in the field 6DFOF search.
      The parameters that define the IHSC are therefore the 
      linking lengths used for the field 3DFOF and 6DFOF search.
      Because galaxies and tidal features are identified using a phase
      space FOF algorithm, the IHSC is composed of
      particles that are too far in phase space to be linked to
      any structure.
      Physically, this means that the component is diffuse and
      kinematically hot.
      This definition allows us to estimate robustly the IHSC for
      systems of different masses in a cosmological simulation.
      {\sc velociraptor} has the ability to recover streams and
      other tidal features, but we do not include these as part of
      the  IHSC; if desired it is straightforward to include those
      structures into the IHSC.
      The mass in these structures is typically less than a few
      per cent the IHSC mass.
      %

      Moreover, the outer extent of the background particles 
      that compose the IHSC can be defined either by the region
      delimited by the 3DFOF field search, or by the virial radius
      $R_{200c}$.
      While the second approach provides a theoretical 
      point of comparison with simulations and observations,
      in the case of merging groups and clusters, some of the
      diffuse material between such systems might be
      left out.
      Throughout this paper we use the former definition,
      although we note that both definitions give equivalent results.
      See Appendix~\ref{appndx:fixedapertures} for a comparison
      of the total and IHSC stellar mass content using
      both of these definitions.
      %
    
      In Fig.~\ref{fig:ihscvis}, we give a visual impression of the projected stellar 
      density of all stellar particles
      (first and third row) and the IHSC
      (second and fourth row) of Milky Way-like mass systems,
      as well as galaxy groups and clusters.
      For reference, we show the extent
      of \emph{fixed spherical} apertures of radius 
      $R_\mathrm{IHSC} = \{30, 100\}$ kpc commonly used to
      separate the central galaxy from the IHSC
      \citep[e.g.][]{Pillepich2018b}.
      While fixed spherical apertures are a simple 
      way of separating the mass content in the central galaxy
      and the IHSC, they do not necessarily provide a robust
      definition that works for systems displaying a variety
      of mass distributions that depart from a spherical symmetry,
      as well as systems of a wide range of masses.
      For example, a spherical aperture of 30
      kpc might be reasonable for Milky Way-like mass systems to
      separate the stellar content of the central galaxy from that of
      the IHSC (e.g. systems \texttt{A} and \texttt{B} in
      Fig.~\ref{fig:ihscvis}).
      However, for galaxies at the centre of groups and clusters
      this aperture can in fact truncate the outer parts of the galaxy
      (e.g. \texttt{D}, \texttt{E} and \texttt{F}).
      In contrast, a larger aperture, e.g. 100 kpc, could work
      for massive systems, but will be too large for
      smaller galaxies encompassing most of the stellar material 
      within the aperture; this can lead to measurements
      of IHSC mass fraction that are too low as a result
      of the measurement method.
      See also Appendix~\ref{appndx:fixedapertures} for a
      comparison between these different definitions.
      %

      To solve these issues, previous studies have
      implemented adaptive spherical apertures based
      on the central galaxy + IHSC (CG+IHSC) mass distribution;
      for example a sphere with radius proportional to the
      CG+IHSC half-mass radius, $R_\mathrm{IHSC} = \beta\,R_{50}$,
      \citep[e.g. $\beta = 2.0$][]{Pillepich2018b,Elias2018}.
      However, it is not straightforward to know which value of
      $\beta$ should be used to give consistent answers across the 
      mass range covered by hydrodynamical simulations.
      Moreover, for this particular choice of $R_\mathrm{IHSC}$,
      the amount of mass that is assigned to the IHSC is likely to
      depend on the specific distribution of the central galaxy rather
      than on the properties of the IHSC.
      For example, the $R_\mathrm{IHSC}$ of a system with a spiral central
      galaxy will be larger than that of an elliptical galaxy with
      same stellar mass only because of their intrinsic mass distribution
      \citep[e.g.][]{vanderWel2014};
      in this case, a lower mass fraction in the IHSC would be expected
      for spirals compared to ellipticals \emph{because of the method} 
      rather than the properties of the IHSC.
      %

      By defining the IHSC as the kinematically hot component, we
      are able to address all the issues that spherical aperture
      definitions have, as is seen in the IHSC projections of
      Fig.~\ref{fig:ihscvis}.
      Moreover, by having the capability to consistently define the IHSC
      for systems with a wide range of masses, we can
      perform statistical studies of the IHSC in entire
      cosmological simulations, as well as systematically following
      the evolution of the IHSC for both individual systems and 
      complete entire populations.
      %

      For simplicity, throughout this paper we refer to 3DFOF 
      objects as ``systems'', which would typically be composed
      by a central galaxy, its satellite galaxies, and the IHSC.

    \subsection{Calculated properties}
      In this Section we describe how we calculate the properties
      of systems and galaxies used in upcoming sections.

      \subsubsection{Number of satellites}
        The number of satellites of a given system (3DFOF) is the
        count of substructures found by {\sc velociraptor}
        composed of more than 50 stellar particles.
        %
        %

      \subsubsection{Mass content}
      The stellar mass of objects will be denoted by the 
      subscript `$*$' and a label.
      This quantity refers to the sum of the mass of all
      stellar particles in that object/component as identified
      by {\sc velociraptor}.
      For the total mass content inside $R_{200c}$, i.e. stars +
      gas + BH + dark matter, we will simply refer to as $M_{200c}$,
      and $M_\mathrm{*,200c}$ for the stellar mass only within
      $R_{200c}$.

      \subsubsection{$V/\sigma$}
      For the kinematic morphology parameter $V/\sigma$ of the
      central galaxies, we first calculate the spherical radius
      that encloses 50\% of the stellar mass of the central
      galaxy\footnote{Note that we consider the central galaxy
      and the IHSC as separate components, therefore no IHSC particles are included in the computation of
      the half-mass radius.}, $R_{50}$.
      Then, we calculate the specific angular momentum, $\mathbf{j}$,
      for all the particles within $R_{50}$, i.e.

      \begin{equation}
          \mathbf{j} = \frac{1}{M_{*,R_{50}}}\sum_i^{r_i \leq R_{50}}
          \mathbf{r}_i\,\times\,m_i \mathbf{v}_i\,,
      \end{equation}
      
      \noindent where $\mathbf{r}_i$ and $\textbf{v}_i$ are the position
      and velocity vectors of stellar particle $i$, with mass $m_i$,
      measured with respect to the galaxy's centre-of-mass; and 
      $M_{*,R_{50}}$ is the mass inside $R_{50}$, or equivalently
      $M_{*,R_{50}} = 0.5\,\mctrl$.
      The rotational velocity $V$ is calculated as

      \begin{equation}
          V = \frac{|\,\mathbf{j}\,|}{R_{50}}\,\,.
      \end{equation}
      
      The velocity dispersion $\sigma$ is calculated as the average 
      3D velocity dispersion of all stellar particles within 
      $R_{50}$,

      \begin{align}
          &\sigma_{v_x}^2 = \frac{1}{M_{*,R_{50}}}\sum_i^{r_i \leq R_{50}} m_i\,v_{x,i}^2\,, \\
          &\sigma = \frac{1}{3}\sqrt{\sigma_{v_x}^2 + \sigma_{v_y}^2 + \sigma_{v_z}^2}\,,
      \end{align}
      
      \noindent where $v_{x,i}$ is the velocity of particle $i$ along
      the Cartesian axis $x$, and $\sigma_{v_x}$, $\sigma_{v_y}$, and 
      $\sigma_{v_z}$ are the velocity dispersions along the $x$, $y$, 
      and $z$ axes, respectively.
      Finally, the kinematic morphology parameter $V/\sigma$ is simply the
      ratio between these quantities.

      \subsubsection{Star formation rate}
      The star formation rate (SFR) of a galaxy is calculated by
      summing the mass of all its stellar particles with age,
      $t_\mathrm{age}$, smaller than a given $\Delta\,t$ window from
      the time of the snapshot analysed, 
      
      \begin{equation}
          \mathrm{SFR} = \frac{1}{\Delta\,t} 
                         \sum_i^{t_{\mathrm{age},i}\,< \Delta\,t}
                         m_{i}\,.
      \end{equation}
      
      \noindent Throughout this paper we adopt a $\Delta\,t = 50$ Myrs
      to calculate this quantity. 
      SFRs calculated using windows of  $\Delta\,t = \{20, 100\}$ Myrs, 
      are consistent with that of $\Delta\,t = 50$ Myrs.
      Mass loss due to stellar evolution is neglected, however;
      if included, it would represent a systematic shift of the
      estimated SFRs, and would not change our conclusions.

  \section{Present day $\mfsm$ relation}
  \label{sec:fmihsc}
    
    The IHSC mass fraction, $\fmihsc$, is calculated as
    
    \begin{equation}
      \label{eq:fmihsc}
      \fmihsc = \frac{\mihsc}{\mstot} \, ,
    \end{equation}
    
    \noindent where $\mihsc$ is the stellar mass of the IHSC, 
    and $\mstot$ is the total stellar mass of the system that
    includes all the stellar mass in the galaxies (central and 
    satellites) and the IHSC.
    As shown in \citet{Canas2019}, {\sc velociraptor}'s galaxy finding
    algorithm is also capable of identifying tidal structures, such as stellar streams and shells, as 
    separate objects that are distinct in phase-space
    from the galaxies and are kinematically colder than the IHSC
    envelope.
    These structures are often incorporated into the IHSC
    because of their diffuse and accreted nature, in the sense that
    their stars do not form in the
    main galaxy/dark matter halo branch, similar to the outskirts of
    galaxies and clusters 
    \citep[\emph{in-situ} vs. \emph{ex-situ}, e.g.][]{Pillepich2015,Dubois2016}.
    We argue, however, that it is important to separate such structures
    from the IHSC as some of that material may be accreted onto the
    galaxy at later times.
    Although these tidal structures can be found with ease by 
    {\sc velociraptor}, for the purposes of this study we consider all
    kinematically distinct stellar structures as satellite galaxies,
    while the IHSC is only the \emph{diffuse} stellar background.

    \begin{figure}
    \includegraphics[width=\columnwidth]{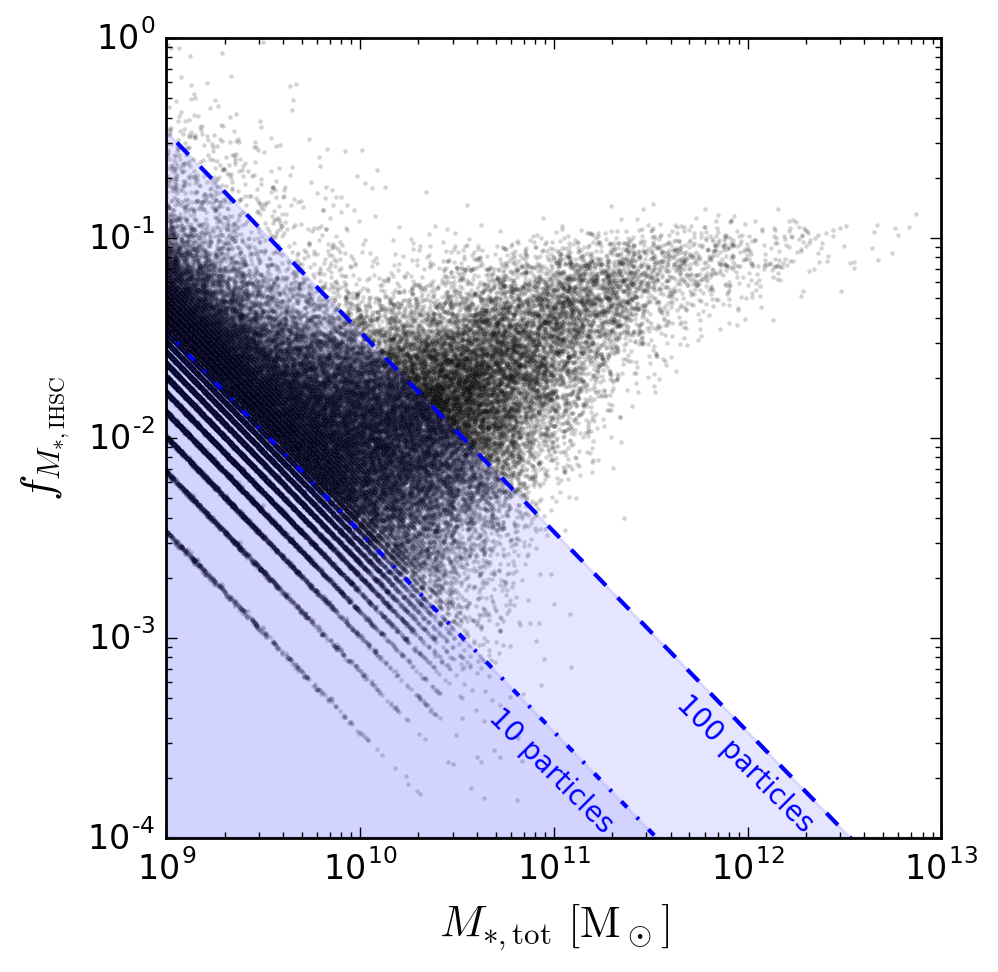}
    \caption{Mass fraction in the Intra-Halo Stellar Component (IHSC), 
             $\fmihsc$, as identified by {\sc velociraptor},
             as a function of total stellar mass $\mstot$ at
             $z = 0$ for all systems in the Horizon-AGN simulation, shown 
             as black dots.
             Blue dashed and dot-dashed lines delimit regions where the
             IHSC at a given $\mstot$ is composed by $\leq 100$
             and $\leq 10$ particles, respectively.
             See text for more details.
             }
    \label{fig:fmihsc}
    \end{figure}
    
    In Fig.~\ref{fig:fmihsc} we show the $\mfsm$ relation in
    Horizon-AGN at $z = 0$.
    The measurements for individual systems (i.e. entire 3DFOF objects)
    are shown as black dots.
    Dashed and dotted-dashed blue lines delimit regions where, 
    at fixed total stellar mass, the IHSC is composed
    of $\leq 100$ and $\leq 10$ particles, respectively.
    At $10^{10.5} \lesssim M_{*,\mathrm{tot}}/\msuni \lesssim 10^{12}\,\msuni$
    $\fmihsc$ increases with increasing $\mstot$,
    with the scatter decreasing as $M_{*,\mathrm{tot}}$ increases.
    At $M_{*,\mathrm{tot}} > 10^{12}\,\msuni$, the relation flattens
    while the scatter keeps decreasing out to the most massive system
    resolved in Horizon-AGN, $M_{*,\mathrm{tot}}\sim 10^{13}\,\msuni$.
    %
    
    The variation in $f_{M_{*,\mathrm{IHSC}}}$ at fixed 
    $M_{*,\mathrm{tot}}$ is $\sim2$ dex at 
    $M_{*,\mathrm{tot}} \simeq 3 \times 10^{10}\,\msuni$
    ($\mvir \simeq 10^{12}\,\msuni$), and gradually decreases towards
    higher $M_{*,\mathrm{tot}}$.
    This is caused by changes in the principle mode of growth
    of a system.
    At $M_{*,\mathrm{tot}} < 10^{12}\,\msuni$ ($\mvir < 10^{13}\,\msuni$)
    systems grow either via continuous infall of gas 
    and star formation; accretion
    of multiple satellite galaxies; or a combination of these two.
    The large dispersion observed at these masses is therefore a consequence
    of the various growth mechanisms these systems can undergo
    (this is demonstrated in Section~\ref{sec:ihscevol}).
    In contrast, mass growth at $M_{*,\mathrm{tot}} > 10^{12}\,\msuni$ 
    is expected to be dominated by interactions and galaxy
    mergers \citep[][]{Oser2010,RodriguezGomez2016,Dubois2013,Dubois2016}.
    Because this is the dominant growth mechanism, we expect that these systems will
    display a tighter distribution in the $\mfsm$ relation.
    It is important to note, however, that the limited volume of 
    Horizon-AGN means that it contains only a handful of systems with $M_{*,\mathrm{tot}} > 10^{12}\,\msuni$,
    and so the tight distribution observed at these masses 
    may be due to poor statistics.
    %
    
    At $\mstot < 10^{10}\,\msuni$ there is an apparent
    anti-correlation between $\mstot$ and
    $\fmihsc$.
    This is a consequence of the resolution limit of the 
    simulation.
    In Horizon-AGN, stellar particles have a mass resolution of 
    $M_* \simeq 3\times10^{6}\,\msuni$, which means that in a galaxy
    composed by 300 particles (i.e. $M_* = 10^{9}\,\msuni$), 
    a $\fmihsc = 0.01$ will be composed by only 3
    particles.
    Such features are visible as diagonal patterns at 
    $\fmihsc < 0.01$ and 
    $\mstot < 10^{10} \msuni$,
    which extends to the upturn observed in the relation and is 
    delimited by the 100 particle limit (blue dashed line).
    This upturn towards lower $M_{*,\mathrm{tot}}$ is also an effect
    caused by the method used: 
    at spatial scales close to the resolution limit of the simulation,
    3DFOF objects can often be decomposed into sparse
    particle distributions with inter-particle distances comparable to
    the 3DFOF linking length;
    by construction such particles are not linked to galaxies by the
    6DFOF search, and are therefore assigned to the IHSC, giving high
    $\fmihsc$ for 3DFOF objects close to the resolution
    limit.
    %

    The overall shape of the relation
    is similar to the \emph{accreted} stellar mass fraction total
    stellar mass relation \citep[e.g.][]{Cooper2013,RodriguezGomez2016},
    but it is important to note that these two quantities are not the same.
    We expect this similarity because the mass budget in the outskirts
    of galaxies and clusters, included in the IHSC, is dominated by
    stars that originated in satellite galaxies, i.e. were accreted
    into the host galaxy/dark matter halo.
    However, accreted mass estimates also take into account the
    stellar mass bound to satellites, as well as stars in the central
    galaxy that were not formed along the main galaxy branch, and is the reason
    why the amplitude of accreted stellar mass fractions is larger
    than $\fmihsc$ at similar $\mstot$.

    \subsection{Parameter dependence of the IHSC}
    \label{sec:paramdep}
      We mentioned in Section~\ref{sec:galihscid} that the IHSC is determined
      by the choice of linking lengths used for the 6DFOF field search.
      As the velocity linking length in the 6DFOF search is chosen to be
      proportional to the velocity dispersion of each system, the only
      user-defined parameter is the configuration space linking length,
      $l_{x,\mathrm{6D}}$.
      This parameter sets a phase-space density cut on the particles
      that are linked to galaxies, excising them from 
      the IHSC.
      Therefore, larger values of $l_{x\mathrm{,6D}}$ will assign more 
      mass to the galaxies and less to the IHSC and vice-versa, as 
      was qualitatively explored in appendix A of \citet{Canas2019}.
      Here, we study in detail the effects different choices of
      $l_{x\mathrm{,6D}}$ have on the estimated IHSC mass,
      $M_{*,\mathrm{IHSC}}$.
      Recalling step 2 of the galaxy identification method used in
      \citet[][section 3.1.2]{Canas2019}, the configuration space
      linking lengths are defined as

      \begin{align}
        \label{eq:lx3d6d}
        \begin{split}
          l_{x,\mathrm{3D}} &= b\ \Delta x \ ,\\
          l_{x,\mathrm{6D}} &= f_{l_{x,\mathrm{6D}}}\ l_{x,\mathrm{3D}}\ ,
        \end{split}
      \end{align}
      
      \vspace{1mm}
      \noindent where $\Delta x$ is the mean inter-particle spacing
      of dark matter particles in the simulation, and $b$ and 
      $f_{l_{x,\mathrm{6D}}}$ are constants $< 1$;
      we adopt the commonly used value of $b = 0.2$.

      \begin{figure}
        \includegraphics[width=\columnwidth]{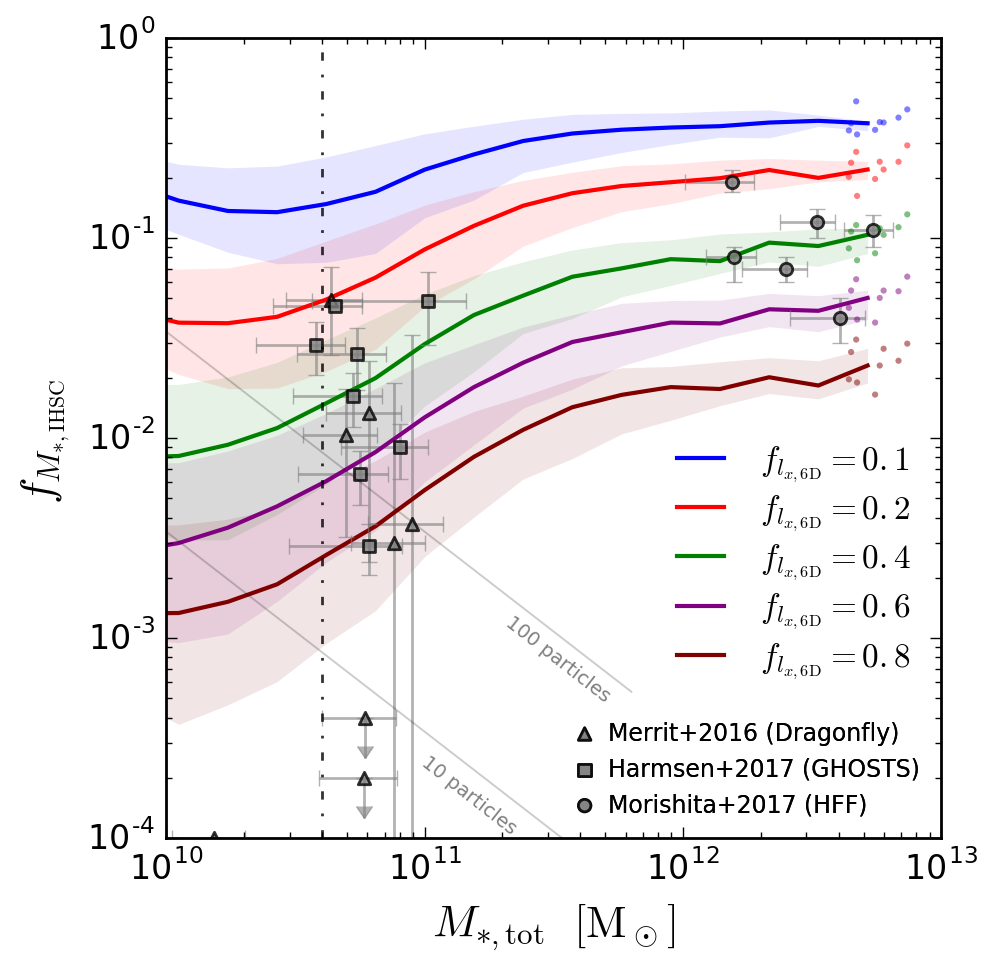}
        \caption{Mass fraction in the IHSC $f_{M_{*,\mathrm{IHSC}}}$ (top left),
                 and mass content in the IHSC and central as estimated using
                 $f_{l_{x,\mathrm{6D}}} = \{0.1, 0.2, 0.4, 0.6, 0.8\}$,
                 coloured as labelled.
                 Solid lines represent the median $f_{M_{*,\mathrm{IHSC}}}$
                 for each $M_{*,\mathrm{tot}}$ bin and shaded regions delimit
                 the 16$^{\mathrm{th}}$ and 84$^{\mathrm{th}}$ percentile
                 of the distribution in that bin;
                 vertical dot-dashed line indicate the $M_{*,\mathrm{tot}}$ 
                 where measurements are considered not to be affected by
                 resolution.
                 Observational estimates of mass fraction in stellar halos
                 from \citep{Merritt2016,Harmsen2017} and ICL 
                 \citep{Morishita2017} are shown as symbols, as labelled.
                 }
        \label{fig:fmihsc_lx}
      \end{figure}

      \begin{figure*}
        \includegraphics[width=0.48\textwidth]{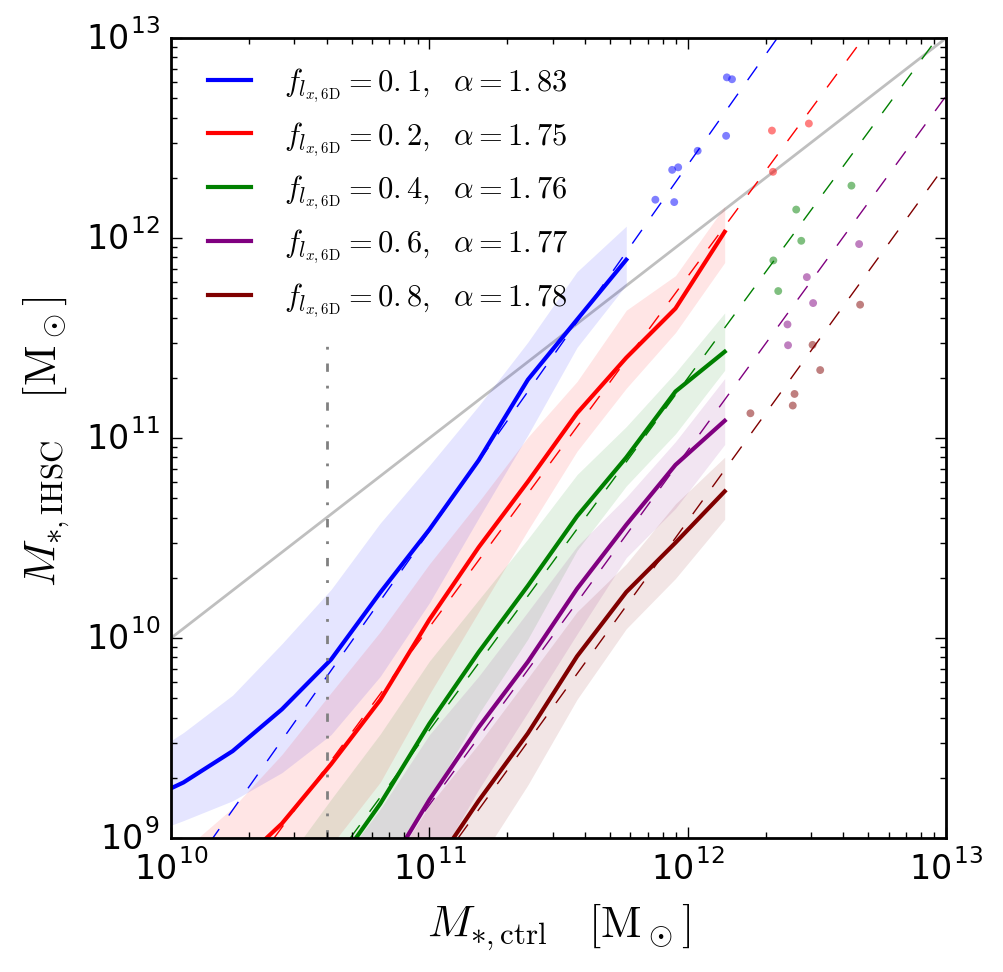}
        \includegraphics[width=0.48\textwidth]{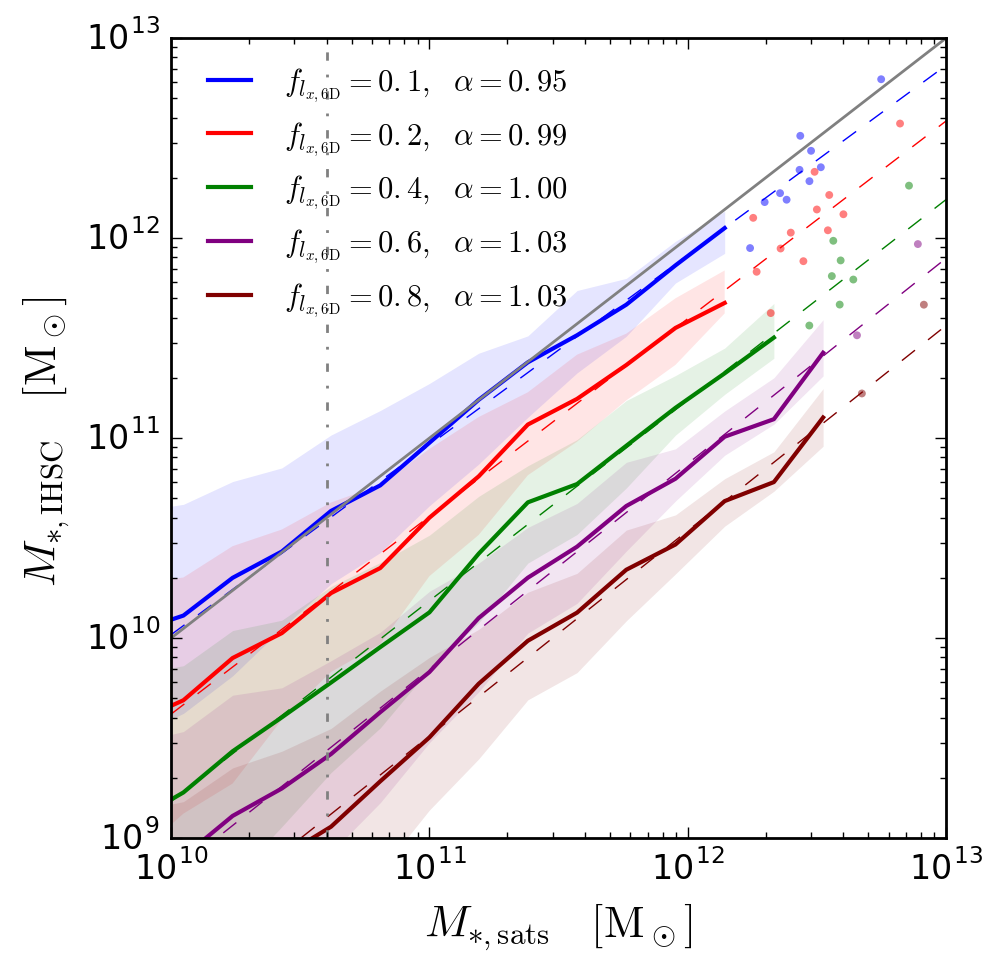}
        \caption{Stellar mass content of the IHSC as a function of the central
                 galaxy stellar mass (left panel), and the total stellar mass
                 in satellite galaxies (right panel) as estimated using
                 $f_{l_{x,\mathrm{6D}}} = \{0.1, 0.2, 0.4, 0.6, 0.8\}$,
                 as labelled.
                 Solid lines represent the median $f_{M_{*,\mathrm{IHSC}}}$
                 for each $M_{*,\mathrm{tot}}$ bin and shaded regions delimit
                 the 16$^{\mathrm{th}}$ and 84$^{\mathrm{th}}$ percentiles.
                 Dashed lines show power-law fit to the \emph{mean}
                 $M_{*,\mathrm{IHSC}}$ (power-law index $\alpha$
                 as labelled) in equally log-spaced resolved $M_*$ bins;
                 bins with less than 5 measurements are not considered,
                 systems in such bins are shown as dots; diagonal gray lines
                 show a 1-to-1 correspondence.
                 These relations are remarkably consistent and independent
                 of the specific $f_{l_{x,\mathrm{6D}}}$ used.
                 The latter demonstrates the robustness of our method to
                 define the IHSC and sets strong predictions
                 that can be easily tested against observations.
                 }
        \label{fig:mihsc_lx}
      \end{figure*}

      In Fig.~\ref{fig:fmihsc_lx} we show the effect that changing
      $f_{l_{x,\mathrm{6D}}}$ has on the $\mfsm$ relation,
      as estimated using different values of $f_{l_{x,\mathrm{6D}}}$;
      solid lines show medians in equal log-spaced mass bins, and
      shaded regions delimit the 16$^{\mathrm{th}}$ and 
      84$^{\mathrm{th}}$ percentiles of the distribution in each mass
      bin.
      Increasing $f_{l_{x,\mathrm{6D}}}$ has the effect of changing the
      overall normalization of the $\mfsm$ relation, in a way that
      increasing $f_{l_{x,\mathrm{6D}}}$ leads to a lower 
      $f_{M_{*,\mathrm{IHSC}}}$.
      In addition, the scatter of the relation below $10^{12}\,\msuni$
      also depends on $f_{l_{x,\mathrm{6D}}}$, with higher values
      leading to larger scatter.
      Despite these differences, the overall behaviour of the 
      $\mfsm$ relation is qualitatively the same: $f_{M_{*,\mathrm{IHSC}}}$
      increases with $M_{*,\mathrm{tot}}$ up to 
      $M_{*,\mathrm{tot}} \sim 10^{12}\,\msuni$, followed by a flattening at
      higher masses.
      The scatter also decreases monotonically with increasing 
      $M_{*,\mathrm{tot}}$, which is in part caused by the limited
      volume of Horizon-AGN.
      The increasing $f_{M_{*,\mathrm{IHSC}}}$ at fixed mass with 
      decreasing linking length is expected as higher $f_{l_{x,\mathrm{6D}}}$
      values result in assigning more particles to galaxies, reducing the
      mass in the IHSC.
      It should be noted that for small $f_{l_{x,\mathrm{6D}}}$ values, 
      resolution effects (i.e. upturn in the relation) become important
      at smaller $M_{*,\mathrm{tot}}$ (vertical dot-dashed line).
      Observations at this stage are shown for reference.

        \begin{figure}
          \includegraphics[width=\columnwidth]{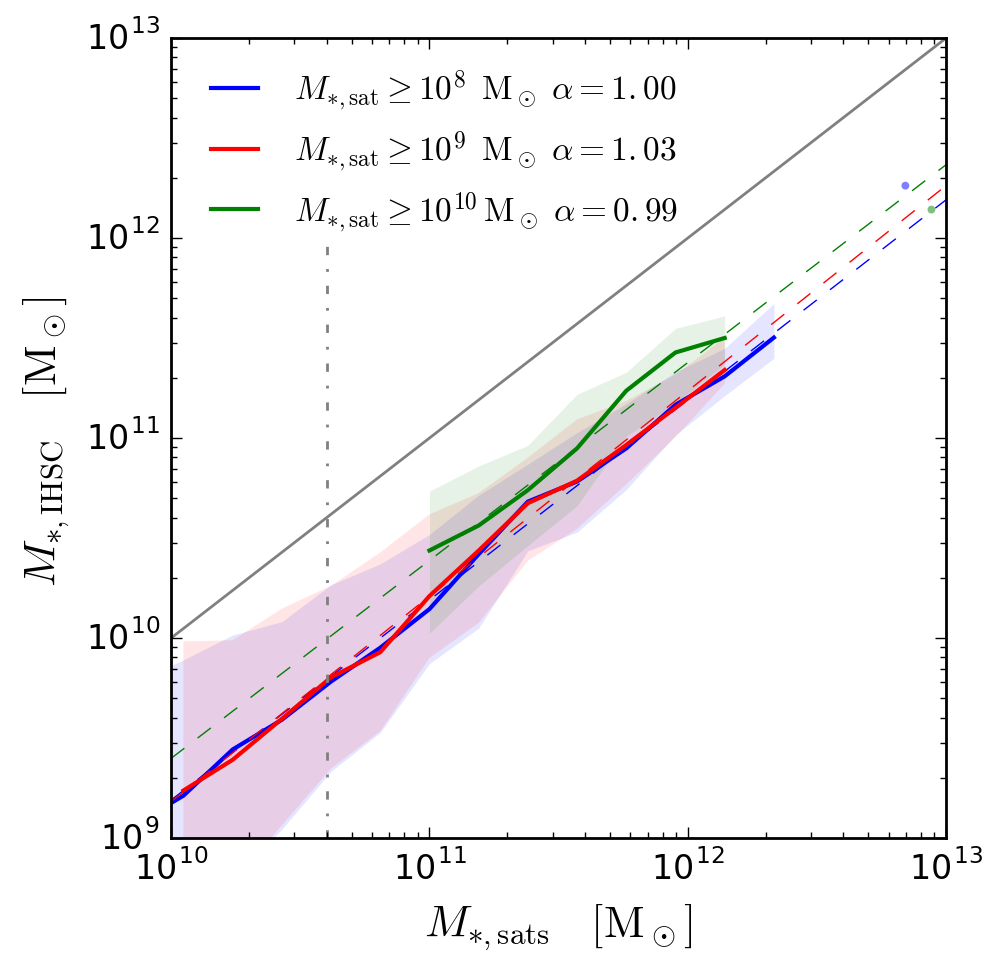}
          \caption{IHSC mass as a function of the total stellar mass
                   in satellites with stellar masses $> 10^{8}$, $10^{9}$, 
                   and $10^{10}\,\msuni$, for $f_{l_{x,\mathrm{6D}}} = 0.4$.
                   Solid lines shows the median in each mass bin, and 
                   shaded regions delimit the 16$^{\mathrm{th}}$ and
                   84$^{\mathrm{th}}$ percentiles.
                   Similarly to Fig.~\ref{fig:mihsc_lx} dashed lines show
                   a power-law fit to the median.
                   The mass relation between $\msats$ and $\mihsc$ is 
                   consistent even when only satellites with $M_* > 10^{8}$
                   and $10^{9}\,\msuni$ are used.
                   For a mass threshold of $10^{10}\,\msuni$ the relation 
                   has a similar slope but is shifted towards the left.
                   }
          \label{fig:mihsc_msats}
        \end{figure}

      In Fig.~\ref{fig:mihsc_lx} we compare the estimated mass in the 
      IHSC to that in the central galaxy, $\mctrl$, and to the total
      stellar mass in satellites, $\msats$;
      For each $f_{l_{x,\mathrm{6D}}}$, we fit a power-law as
      $M_{*,\mathrm{IHSC}} \propto M_{*}^\alpha$,
      to the mean $M_{*,\mathrm{IHSC}}$ per $M_{*}$ bin, 
      shown as dashed coloured lines, with power-law index
      $\alpha$ as labelled;
      bins are required to have at least 5 systems and a
      $M_{*,\mathrm{tot}} > 4\times10^{10}\,\msuni$ to avoid
      consideration of systems affected by resolution;
      points show individual measurements of resolved systems
      in bins with less than 5 systems.
      The mass relations between these components are strikingly
      similar for different values of $f_{l_{x,\mathrm{6D}}}$, with 
      power-law index variations smaller than 5\%. 
      This agreement is remarkable due to the large range of 
      $f_{l_{x,\mathrm{6D}}}$ and stellar masses explored.
      Moreover, the extrapolation of the fit to each distribution
      does seem to be in agreement as well with mass estimations of 
      high mass systems where statistics are low.
      %
      
      The agreement between the mass relations shows that the
      robustness of our results rely on the methodology and not on
      a specific choice of 
      $f_{l_{x,\mathrm{6D}}}$\footnote{We must remember that as with 
      any other method, poorly chosen parameters affect
      properties of individual galaxies, e.g. extremely short values
      of $f_{l_{x,\mathrm{6D}}}$ are likely to 
      identify only the cores of galaxies.
      See appendix A of \citet{Canas2019} for a visual example.}.
      This consistency can also be used to make predictions
      that can easily be tested by observations.
      For example, the fact that $\alpha\sim1$ for the $\mihsc-\msats$
      relation tells us that \emph{on average} the IHSC mass is
      a constant fraction of the total mass in satellites.
      In fact, although in Fig.~\ref{fig:mihsc_lx} $\msats$ includes
      all substructures found by {\sc velociraptor} composed of at
      least 50 particles (i.e. $M_* \geq 10^{7}\,\msuni$), the power-law
      index $\alpha$ is also consistent if mass thresholds of 
      $10^{8}$ and $10^{9}\,\msuni$  are used, as shown in 
      Fig.~\ref{fig:mihsc_msats}.
      These results highlight the advantage of using phase-space
      iso-density cuts over fixed spherical aperture definitions of
      the IHSC, where the choice of parameter $R_\mathrm{IHSC}$ does
      affect the slope measured \citep[e.g.][]{Pillepich2018b}.
      The latter can be understood because the stellar mass
      distributions of systems of different mass (e.g. MW-like 
      and cluster) are not self-similar, nor necessarily 
      spherically symmetric;
      therefore, the mass enclosed within spherical shells 
      assigned to the central (IHSC) for an increasing (a decreasing)
      $R_\mathrm{IHSC}$ correspond to a varying fraction across 
      $\mstot$, and consequently they will not preserve the mass relation
      between these components.
      See Appendix~\ref{appndx:fixedapertures} for complementary
      comparisons to spherical apertures.

    \begin{figure}
      \includegraphics[width=\columnwidth]{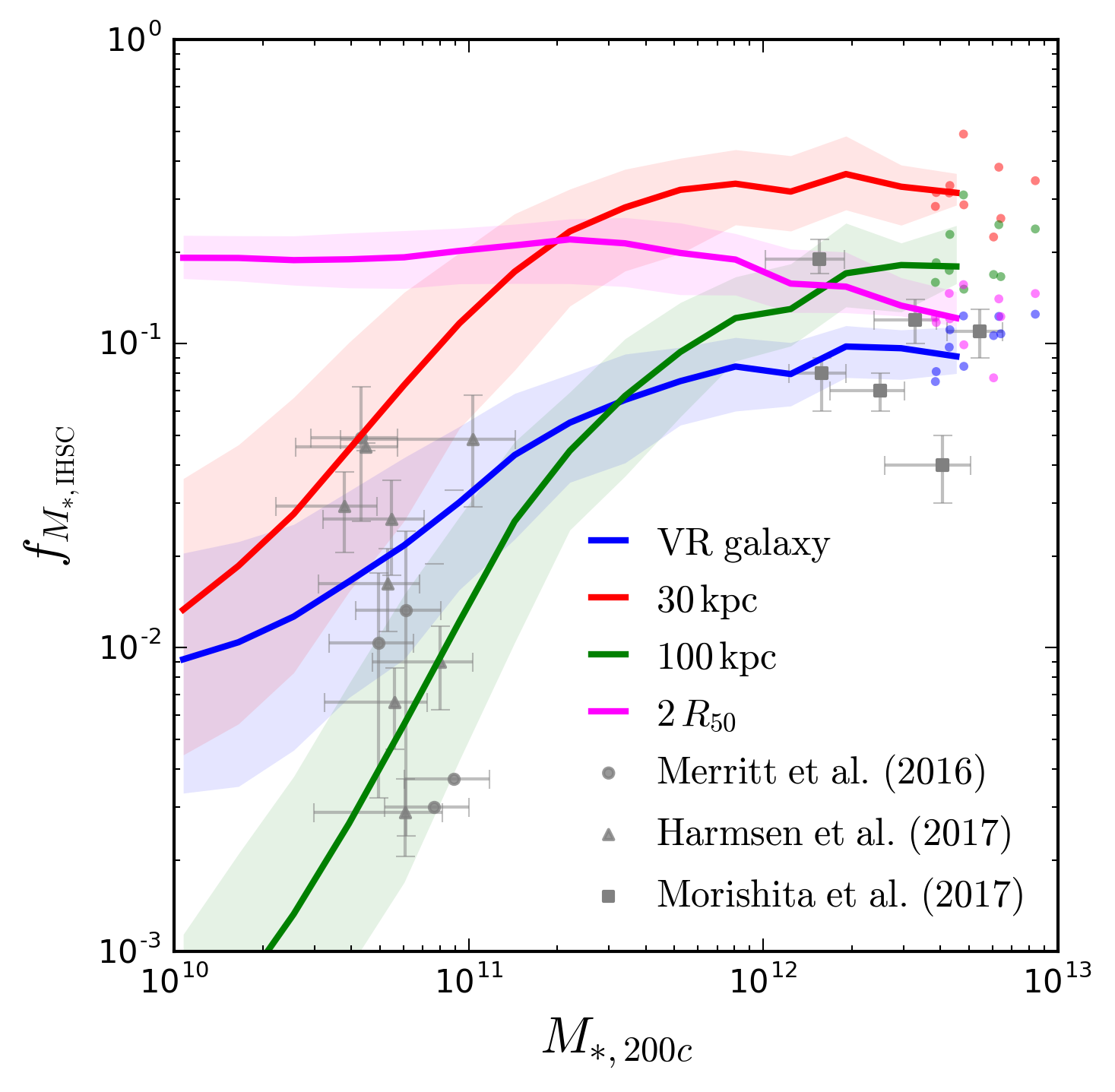}
      \caption{Mass fraction in the IHSC, $\fmihsc$, as a function of
               total stellar mass within spherical overdensity $M_{*,200c}$
               (top) and total halo mass $M_{200c}$ (bottom).
               At $M_{*,200c} < 10^{11}\,\msuni$ compared to observations
               $2\,R_{50}$ overestimates the estimated $\fmihsc$, 100 kpc
               aperture underestimates it, and {\sc velociraptor} and 30 kpc
               aperture are in better agreement.
               At $M_{*,200c} > 10^{12}\,\msuni$, fixed apertures
               overestimate the $\fmihsc$ estimated by observations, while
               a better agreement is shown by {\sc velociraptor} and
               $2\,R_{50}$ aperture.
               Our adaptive phase space method allows us to get IHSC mass
               fractions in better agreement with observations in the whole
               mass range.
               }
      \label{fig:fmihsc_ms200c}
    \end{figure}

      The observed behaviour of our method is quite a useful 
      finding for both simulations and observations because 
      definition of the IHSC (i.e. stellar halo, IGL, ICL) often relies
      on defining or delimiting the extent of galaxies, which is
      not a straightforward task because there is no `true' definition
      of where galaxies end.
      In this study we adopt $f_{l_{x,\mathrm{6D}}} = 0.4$, because it
      predicts a $\mfsm$ relation that is in better agreement with 
      the estimated stellar halo mass fractions
      for Milky Way-like galaxies from \citet{Merritt2016} and
      \citet{Harmsen2017}, and recent ICL mass estimations
      of \citet{Morishita2017}.
      Although the exact method to measure the stellar halo and ICL
      mass fraction differs between studies, these measurements
      give a rough idea of the expected mass fraction in different mass
      ranges; this provides us with an equivalent phase-space 
      density threshold, which in our method is equivalent to a fixed
      $f_{l_{x,\mathrm{6D}}}$.

    \subsection{Comparison to IHSC definitions in literature}
    \label{sec:compsph}
    As mentioned above, for simplicity and ease of comparison,
    studies in the literature have used geometrical definitions
    of the IHSC.
    In Fig.~\ref{fig:fmihsc_ms200c}, we compare the $\fmihsc$
    obtained with our method to that of spherical apertures with
    $R_\mathrm{IHSC} = 30\,\mathrm{kpc},\,100\,\mathrm{kpc},\,2\,R_{50}$.
    To properly compare the methods we show the $\fmihsc$ as
    a function of the total stellar mass within $\rvir$, i.e.
    $M_{*,200c}$.
    A visual representation and a comparison of the mass content
    with these definitions is shown in Appendix~\ref{appndx:fixedapertures}.
    Observations are shown for reference.
    At Milky Way scales, i.e. $M_{*,200c} \sim 10^{11}\,\msuni$, 
    $2\,R_{50}$ overestimates the $\fmihsc$, as the median of their
    distribution goes through the galaxies with the highest $\fmihsc$
    of the observed sample; 
    the 30 kpc aperture median is in slight better agreement, while
    the 100 kpc aperture underestimates the $\fmihsc$ at these
    masses.
    At the high mass both fixed apertures overestimate the $\fmihsc$
    compared to the estimated from the Frontier Field clusters, while
    $2\,R_{50}$ is in better agreement with these measurements.
    Only our method is capable of predicting IHSC mass fractions that
    are in agreement with observations in a wide range of masses because it offers an adaptive, physically motivated, shape-independent and consistent definition of
    the IHSC throughout the mass range, representing a considerable
    advantage over previous studies.

  \section{Unveiling the nature of the $\mfsm$ relation}
  \label{sec:unveiling}
    One of the aims of this study is to understand the origin
    of the scatter observed in the $\mfsm$ relation 
    \citep[e.g.][]{Merritt2016,Harmsen2017}.
    In this section, we explore how the mass fraction in
    the IHSC, $f_{M_{*,\mathrm{IHSC}}}$, correlates with
    observables, specifically with those that relate
    directly to environment and accretion history.
    We remind the reader that throughout this paper we refer
    to a `system' as
    the ensemble of components in a 3DFOF object, composed of
    a central galaxy, satellites, and its IHSC.
    Here, we focus on \emph{static} properties of systems, 
    i.e. properties that can be measured at a fixed 
    time-step, to provide guidance on possible
    third parameters that could be explored in observations
    that would help to interpret the observed scatter.
    All the measurements presented in this Section are
    done at $z = 0$; in subsequent sections we will focus on
    their temporal evolution.

    \subsection{Number of satellites}
\begin{figure} 
        \centering
        \includegraphics[width=\columnwidth]{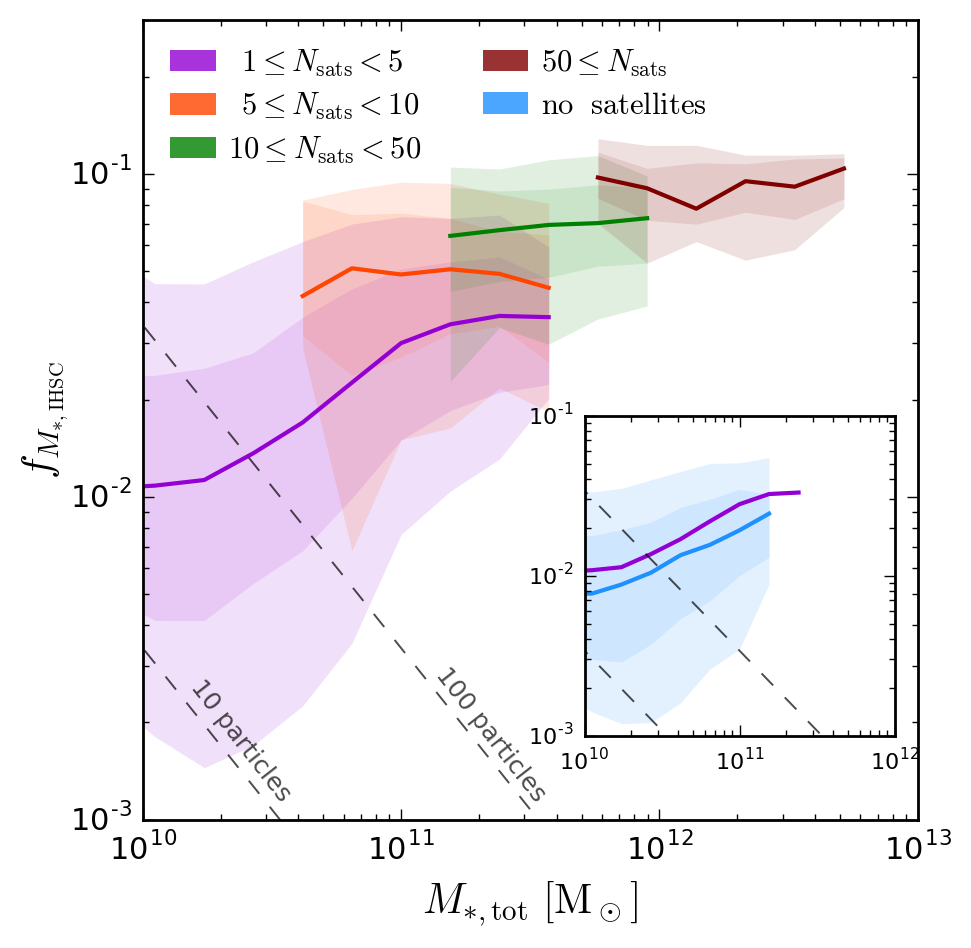}
        \caption{$\mfsm$ relation separating it into sub-samples of 
                 systems with different total number of satellites,
                 $N_\mathrm{sats}$.
                 Solid lines show the median of the distribution, and 
                 shaded regions delimit the $1\sigma$ and $2\sigma$, 
                 respectively.
                 Colours show the sub-samples using different 
                 $N_\mathrm{sats}$ thresholds.
                 Inset in the top panel show the contribution of
                 systems with no identified satellites.
                 }
        \label{fig:mihsc_mstot_nsat}
      \end{figure}

      As the IHSC is built from the tidal debris of orbiting
      satellites and mergers, we expect that the mass in the IHSC should correlate with the accretion history of a halo, as well as with the
      availability of material that can be deposited into the IHSC.
      In Fig.~\ref{fig:mihsc_mstot_nsat} we show the $\mfsm$ relation
      for sub-samples according to the \emph{current} number of
      satellites, $N_\mathrm{sats}$, as identified by {\sc velociraptor}.
      %
      
      Overall, more massive systems tend to have a higher number of 
      satellites, as expected from hierarchical growth.
      However, in the range in which there is overlap between
      the samples with different number of satellites, we see that a
      system with a larger total number of satellites display a higher
      $f_{M_{*,\mathrm{IHSC}}}$ at fixed $M_{*,\mathrm{tot}}$.
      Although the scatter around the $\mfsm$ relation in each
      sub-sample is large, the correlation between $\fmihsc$ and 
      $\nsats$ is noticeable at
      $2\times10^{11} \lesssim \mstot/\msuni \lesssim 4\times10^{11}$,
      in which the distribution of three subsamples of different
      $\nsats$ overlap.
      However, the $1\,\sigma$ regions are typically large enough
      that they overlap, and hence a point in the 
      $\mfsm$ plane cannot be robustly associated with a single
      value of $\nsats$.

      It is important to note that for lower
      masses and less populated systems, the scatter is quite large,
      $\sim 0.5$ dex ($\sim 1$ dex) for $1\sigma$ ($2\sigma$).
      This can be caused by a variety of factors. For example,
      at fixed $\mstot$ a system with only one satellite that has just
      accreted is likely to have a very different $\fmihsc$ than
      another one-satellite system in which the satellite has had enough
      time to interact with the central galaxy, and therefore deposit
      material into the IHSC.
      In fact, after only one orbit, an average subhalo 
      is expected to
      lose $\sim50$\% of its infall mass after its first peri-centric passage, and
      up to $\sim80$\% after its second (Poulton et al. in prep). For this reason, we expect to be within a given sub-sample's scatter in the $\mfsm$ relation.
      Moreover, a system without satellites could have a quiescent
      accretion history, or its satellites may have already
      merged with the central galaxy.
      This possibility can be seen in the inset where the
      distribution of systems without satellites is shown;
      although the overall amplitude of $\fmihsc$ is lower than we see in 
      the $1 < \nsats < 5$ sample, both the medians and
      the size of scatter are comparable, almost overlapping.
      %

      If we repeat the exercise of Fig.~\ref{fig:mihsc_mstot_nsat} but
      limit it to satellites above $10^9\,\msuni$ 
      \citep[which is a typical value adopted to avoid resolution effects; see e.g.][]{Dubois2016}, 
      we find that the relation between $\mfsm$ and $\nsats$ prevails, 
      but becomes weaker.
      If we limit ourselves to only the massive satellites, 
      $M_* > 10^{10}\,\msuni$, we find no dependence of the scatter
      of the $\mfsm$ on $\nsats$.
      This is not necessarily surprising because the dynamical
      friction timescale of massive galaxies is much shorter than
      low mass ones, and hence the current number of massive
      satellites is not necessarily a good indication of the 
      overall assembly history of a system.

    \subsection{Dynamical State}
    \label{sec:dynamicalage}
    
      \begin{figure}
        \includegraphics[width=\columnwidth]{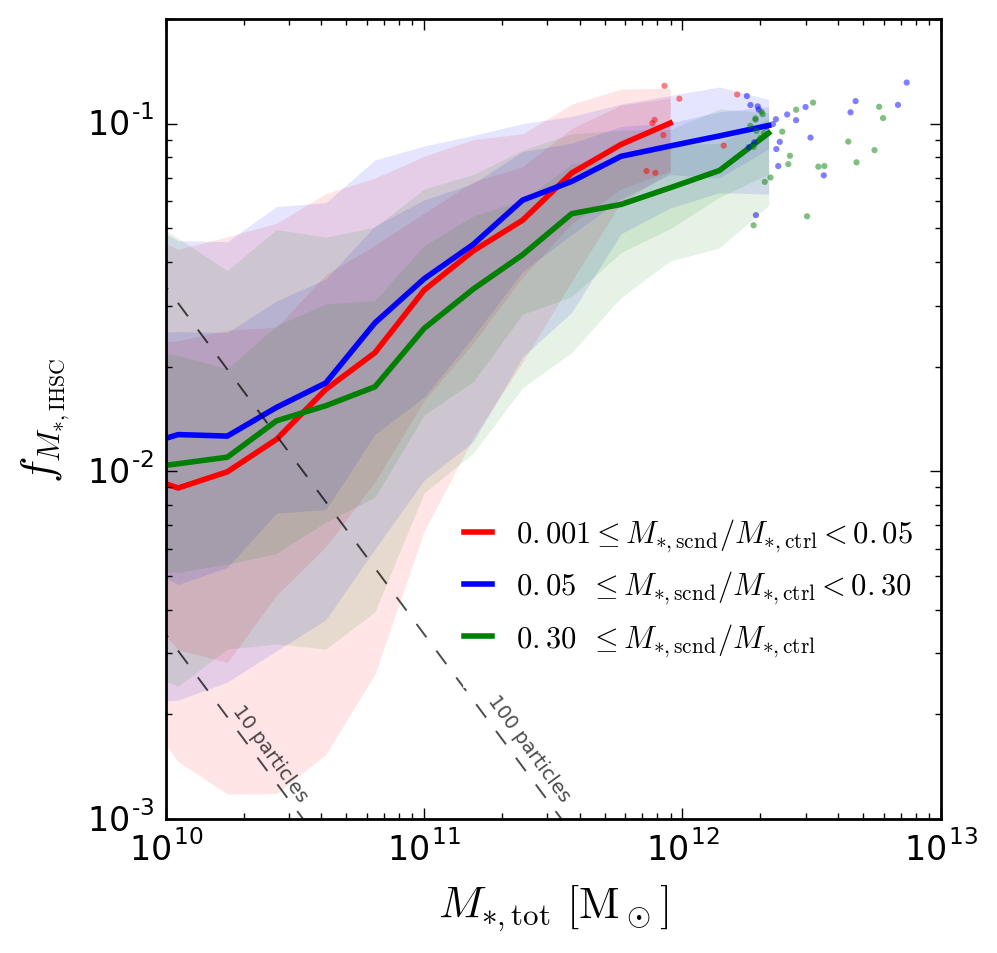}
        \caption{Same as Fig.~\ref{fig:mihsc_mstot_nsat} the $\mfsm$
                 relation into sub-samples of galaxies of different
                 $\mscnd/\mctrl$, as labelled.
                 The overlap between the medians and $\sigma$ contours
                 of different sub-samples, indicates that there is no
                 correlation between $\fmihsc$ and $\mscnd/\mctrl$ for 
                 systems with $\mstot > 10^{11.5}$.
                 At the high-mass end, the distribution of sub-samples
                 suggest that in fact this proxy for dynamical age is 
                 a good indicator of the expected IHSC mass fracton of
                 a system, with dynamically younger systems displaying a
                 lower $\fmihsc$, and vice-versa;
                 the latter being consequence of the dominant growth
                 mechanisms at this mass range.
                 See text for further details.
                 }
        \label{fig:mihsc_fracmscnd}
      \end{figure}

      While the abundance of substructures give information about the 
      environment of a system, there is no evolutionary information that
      can be obtained by a satellite count.
      Because the IHSC is built from disrupted material from ongoing and
      past interactions between galaxies, as mentioned above, the expected
      $\fmihsc$ of a system is likely to be different at the time a
      satellite has 
      just been accreted compared to a later epoch when interactions
      have already taken place and mass has been deposited into the IHSC.
      We explore this idea by using the ratio between the masses
      of the largest satellite (or second most massive galaxy), $\mscnd$,
      and the central galaxy, $\mctrl$, as a proxy for the dynamical
      age of a system.
      This definition is widely used in observations because it is 
      readily accessible \citep[e.g.][]{Davies2019}.
      As dynamical friction is more efficient on massive satellites, 
      their orbits are expected to decay more quickly
      than those of less massive satellites, and they
      will have more signficant interactions with the central galaxy more quickly than less massive 
      substructures; this suggests that a higher $\mscnd/\mctrl$ should
      indicate that the system is young, and debris from 
      interactions are yet to be deposited into the IHSC.
      On the other hand, smaller $\mscnd/\mctrl$ are typically associated
      with dynamically old systems, indicating that interactions with larger
      satellites might have already occurred, and mass has been deposited
      onto the IHSC.
      %
      
      In Fig.~\ref{fig:mihsc_fracmscnd} we show the $\mfsm$
      relation for sub-samples of different $\mscnd/\mctrl$;
      using mass ratios of $[0.001,0.05)$, $[0.05,0.3)$ and
      $[0.3,1.0)$, which mimic commonly adopted
      thresholds to classify mergers as mini, minor and major,
      respectively.
      At $M_{*,\mathrm{tot}} < 10^{11.6}\,\msuni$ we find that the
      three samples of $\mscnd/\mctrl$ have similar medians as
      well as 1 $\sigma$ and 2 $\sigma$ contours.
      It is only at $M_{*,\mathrm{tot}} > 10^{11.6}\,\msuni$ that
      we start to see the expected dependence of a higher
      $f_{M_{*,\mathrm{IHSC}}}$ as $\mscnd/\mctrl$ decreases
      at fixed $\mstot$.
      The fact that $\mscnd/\mctrl$ disentangles 
      the distributions only at larger $\mstot$ is because mergers are the principal mechanisms through 
      which these systems grow
      (see for example \citealt{Oser2010},\citealt{Dubois2016} 
      and \citealt{RodriguezGomez2016} for simulations,
      and \citealt{Robotham2014} for observational evidence).
      While these results are indicative, stronger conclusions
      cannot be made because of the small number of systems with 
      $\mstot > 10^{12}\,\msuni$ that can be resolved by the Horizon-AGN
      simulation box ($L_\mathrm{box} = 100\,h^{-1}\,$ Mpc).

    \subsection{Kinematic Morphology}
      The internal properties of the central galaxy  provide
      additional information about the accretion history of a system.
      Previous studies have found that galaxies
      with low $V/\sigma$ are mostly formed by mergers
      \citep{Dubois2013,Dubois2016}, particularly
      dry mergers \citep{Lagos2018b}.
      Although several routes can lead the formation of the so-called
      slow rotators \citep{Naab2014}, large cosmological hydrodynamical
      simulations have allowed the statistical inference of the dominant
      mechanism behind their formation \citep{Penoyre2017,Lagos2018b}.
      Therefore, we expect that kinematic morphology gives us an 
      indication of the merger history of the galaxy.
      Here, we use the kinematic morphology $V/\sigma$ of the central galaxy
      as an indicator of the accretion history experienced by a system.
      We note that $V/\sigma$ for Horizon-AGN galaxies does indeed
      reflect the properties expected for a galaxy given its
      visual morphology; see for instance figure 2 of 
      \citet{Dubois2016}\footnote{Note that although the exact way of
      calculating $V/\sigma$ for this study differs from that of 
      \citet{Dubois2016}, low (high) $V/\sigma$ values correspond
      to spheroidal (disc) dominated galaxies for both methods.}.
      %
      
\begin{figure}
        \includegraphics[width=\columnwidth]{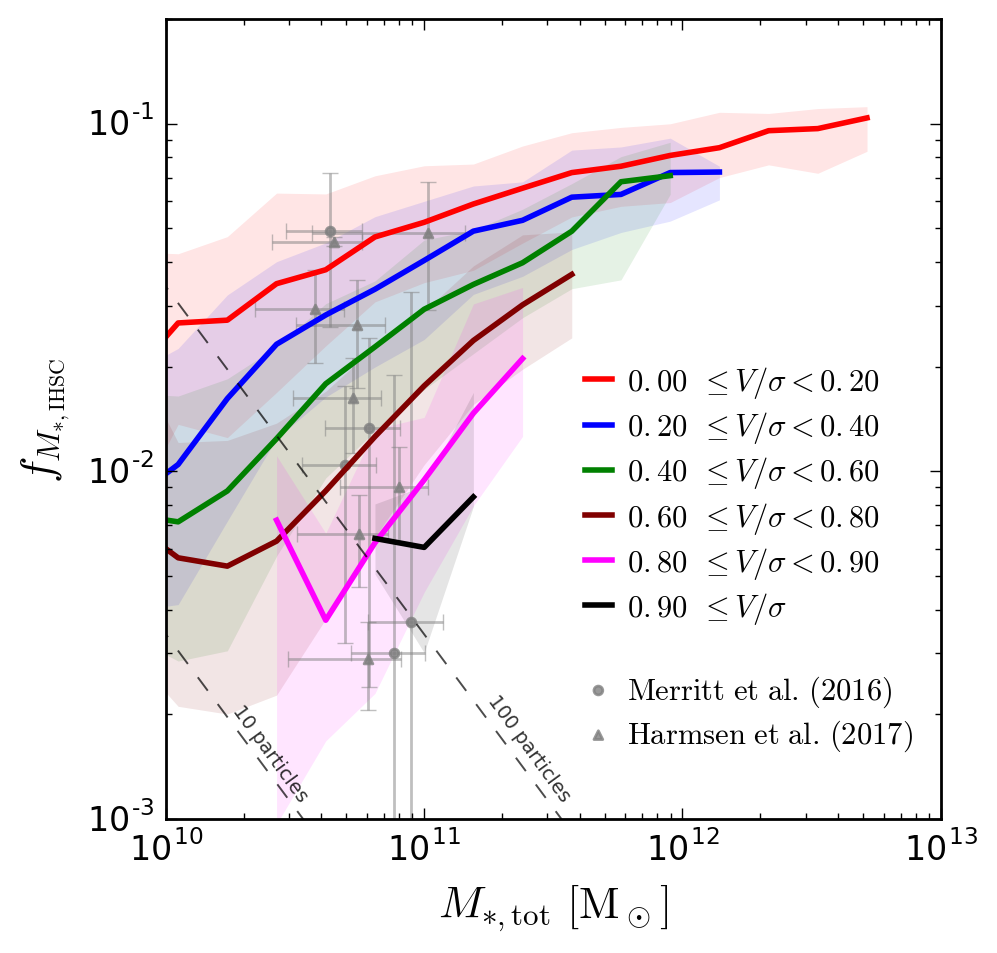}
        \caption{Same as Fig.~\ref{fig:mihsc_mstot_nsat} but the $\mfsm$
                 relation into sub-samples of galaxies of different $V/\sigma$,
                 as labellled.
                 For clarity we only show only 16$^\mathrm{th}$-84$^\mathrm{th}$
                 percentiles shaded regions.
                 Observational estimates from \citet{Merritt2016} and
                 \citet{Harmsen2017} are shown as grey symbols.
                 It is clear that the $\fmihsc$ does show an anti-correlation
                 with the kinematic morphology $V/\sigma$ of the central 
                 galaxy;
                 the medians of each sub-sample do nor overlap and show
                 a continuous transition from high $V/\sigma$ and low
                 $\fmihsc$ towards low $V/\sigma$ and high $\fmihsc$.
                 The observed behaviour is consistent with the expected active 
                 assembly history of galaxies supported mainly by dispersion,
                 and a more quiescent one for rotationally supported ones.
                 See text for a more extensive discussion.
                 }
        \label{fig:mihsc_vsigma}
      \end{figure}

      In Fig.~\ref{fig:mihsc_vsigma} we show the $\mfsm$ relation,
      separating systems into sub-samples according to the
      $V/\sigma$ of their central galaxy, as labelled.
      At all masses, systems with high $\fmihsc$ host a
      highly dispersion-supported galaxy, while a lower $\fmihsc$
      corresponds to systems hosting a rotationally-supported galaxy.
      There is a clear distinction between different sub-samples
      as their medians do not overlap or cross each other;
      moreover, there is a gradual transition between sub-samples
      from low $\fmihsc$ and high $V/\sigma$, to high $\fmihsc$
      and low $V/\sigma$.
      These results are consistent with the notion that  rotationally
      supported galaxies should have had quiescent accretion 
      histories, and consequently lower $\fmihsc$.
      \citet{Lagos2018b} found that the highest $V/\sigma$
      galaxies can only be obtained in the absence of mergers.
      On the other hand, a dispersion-supported galaxy is expected
      to have experienced multiple mergers and interactions, which in
      turn should deposit higher amounts of mass into the IHSC,
      increasing its $\fmihsc$.

      We find that the strongest correlation displayed
      by the kinematic morphology of the central galaxy 
      is with the scatter of the $\mfsm$ relation at 
      $\mstot < 10^{12}\,\msuni$.
      This may be surprising because other parameters studied here are
      typically used in observations to separate dynamically young and old systems.
      This means that, in Horizon-AGN, kinematic 
      morphology is the
      best indicator of how the IHSC was built up.
      Previous simulation results suggest that the IHSC could be
      easily built up by dry mergers, because the mass of satellites is 
      preferentially deposited in the outskirts 
      \citep[see e.g.][]{Lagos2018a}.
      This, combined with the fact that simulations predict dry mergers
      to be the most effective way of producing slow rotators, may be
      the reason behind our findings.
      For higher masses, the dynamical state of the system gives
      us a handle on the scatter in a way that morphology cannot, because
      at $\mstot > 10^{12}\,\msuni$ all the centrals are
      mainly supported by dispersion.
      Note that there is a difference between the $V/\sigma$
      calculated (which uses the 3D velocity information of galaxies)
      with what is recovered from a 2D projection of that information.
      This is partly due to inclination and partly due to limitations
      in resolution in observations that affect the inferred velocity
      dispersion in observations (see \citealt{vandeSande2019} for a
      detailed analysis of how hydrodynamical simulations compare to 
      IFU surveys).
      %
      
      In recent work \citet{Elias2018} found an anti-correlation
      between the mass fraction in the stellar halo and the
      morphology parameter $\kappa$ in the Illustris
      simulation.
      While in principle this is consistent with the anti-correlation
      observed in Fig.~\ref{fig:mihsc_vsigma}, we argue that
      these are not completely equivalent because of the spherical
      aperture of $2\,R_{50}$ (magenta line in
      Fig.~\ref{fig:fmihsc_ms200c}) used to separate the stellar
      halo.
      This definition not only overestimates the IHSC/stellar halo
      at Milky Way masses, but also fails to reproduce the observed
      $\fmihsc$ scatter.
      Consequently the $\fmihsc$ scatter at fixed $\kappa$ is
      comparable to that of the entire population.
      %

      We can, in principle, connect this result with those
      found in recent observations by \citet{Merritt2016} and
      \citet{Harmsen2017} (shown as symbols in 
      Fig.~\ref{fig:mihsc_vsigma}), who, using different observational
      techniques, found that the estimated $\fmihsc$ for Milky
      Way-like galaxies can vary up to 2 dex in a similar mass range.
      Because of their disc morphology, we would expect all the galaxies in
      these studies to have a high $V/\sigma$, and consequently a low
      $\fmihsc$ according to the behaviour observed in our results.
      However, their location in the relation overlaps with the entire
      simulated population, even for simulated galaxies mainly supported
      by dispersion.
      While this can be in conflict with the trend observed in
      our measurements, we have to take into account that the observational
      data do not include ellipsoidal galaxies which are needed to test
      if the $V/\sigma$ correlation is displayed by observations.
      The `tension' can be alleviated by shifting the $\mfsm$ relation
      measured from the simulation towards higher $\fmihsc$ values, which
      can be achieved by using a different choice of $f_{l_{x,\mathrm{6D}}}$,
      as shown in Section~\ref{sec:paramdep}; 
      however, this would then introduce a conflict with estimated mass
      fraction in the ICL, and would also reduce the extent and mass
      of the galaxies (see Section~\ref{sec:paramdep}, and
      appendix A of \citealt{Canas2019}).
      Another factor that comes into play is the uncertainty in the
      measurements as well as the scatter that individual $V/\sigma$
      sub-samples have;
      $1\sigma$ of contours of consecutive $V/\sigma$ sub-samples
      overlap and $2\sigma$ contours as large as $\sim 0.5$ dex, indicating
      that, although unlikely, a galaxy with high $V/\sigma$ can high 
      $\fmihsc$ as well, and vice-versa.

    \subsection{Specific star formation rate}
      We have shown that $V/\sigma$ can be used to estimate the 
      mass fraction in the IHSC.
      However, observational estimates of this parameter 
      require expensive spectroscopic observations, and while
      surveys such as MaNGA \citep{Drory2015}, CALIFA 
      \citep{Sanchez2011}, and SAMI \citep{Croom2012} have
      been capable of estimating kinematic properties of thousands
      of galaxies, there are no estimates of the IHSC
      of these galaxies to date.
      Here we explore a possible correlation
      between specific star formation rate (sSFR) of the central galaxy
      and the $\fmihsc$.
      Star formation rate and $V/\sigma$ of a
      galaxy are strongly correlated, and it is easier to estimate sSFR from observations that will be readily accessible via SED fitting to a survey such as LSST \citep{Robertson2017}.
      %

\begin{figure}
        \includegraphics[width=\columnwidth]{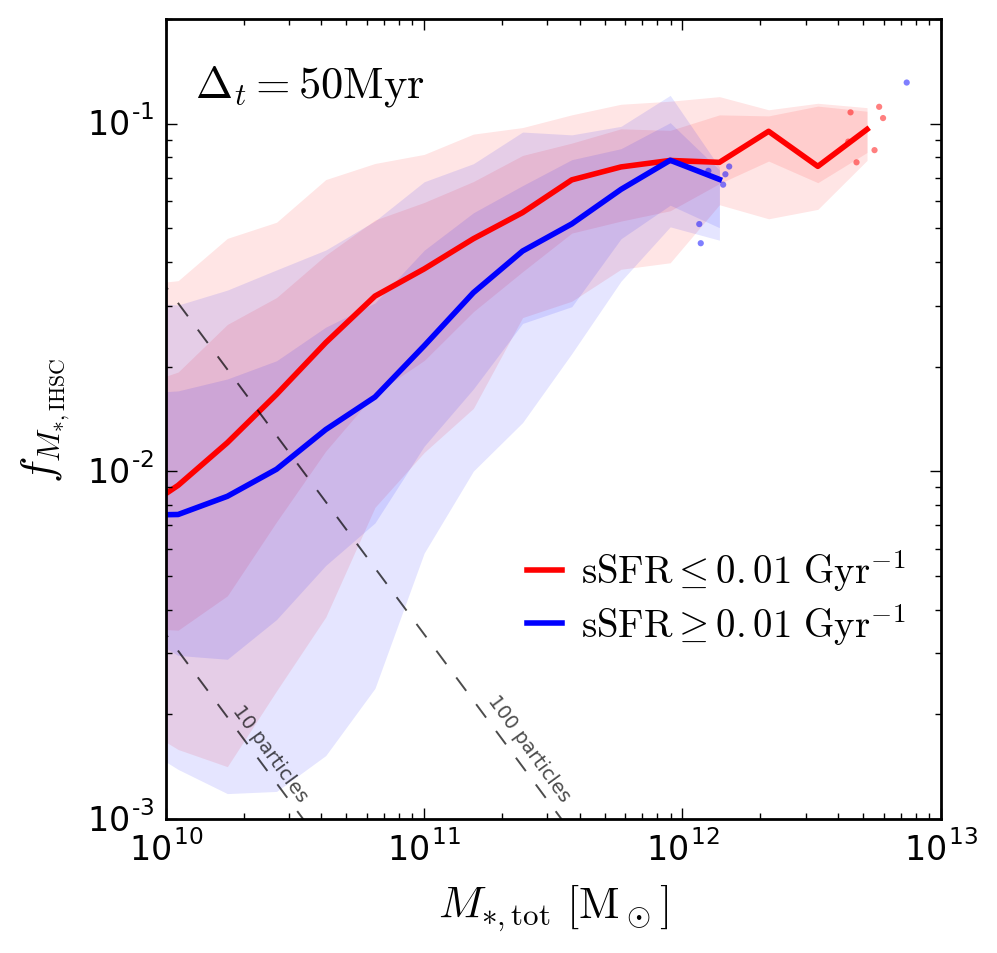}
        \caption{Same as Fig.~\ref{fig:mihsc_mstot_nsat} but
                 separating the $\mfsm$ relation into sub-samples of 
                 galaxies of different sSFRs, as labelled.
                 The sSFR shows a weak trend with $\fmihsc$ compared
                 to $V/\sigma$.
                 There is, however, a slight indication
                 that higher sSFR have lower $\fmihsc$.
                 A higher number of sSFR bins give similar results,
                 for clarity we show only two sub-samples where the
                 relations are more distinct.
                 }
        \label{fig:mihsc_ssfr}
      \end{figure}

      In Fig.~\ref{fig:mihsc_ssfr} we show the $\mfsm$ relation for
      sub-samples of sSFR.
      Although the median of each sub-sample indicates that there 
      is anti-correlation between the sSFR and $\fmihsc$, the scatter 
      observed is quite large.
      A larger number of sub-samples shows a similar behaviour;
      because of highly overlapping distributions, we only show two
      sub-samples that are most distinct for clarity.
      We repeated this calculation using bins in SFR instead of
      sSFR, and found similar results.
      The observed behaviour is a consequence of several factors.
      One is the way in which we estimate the SFR;
      by using only stellar particles to determine the SFR, it is
      likely that our estimate is affected by particle discreteness and mass resolution, because we cannot measure SFR
      below one stellar particle formed in the $\Delta\,t$ used.
      Another is that spheroidal and disk galaxies of similar masses can have similar SFRs; this can arise if star formation quenching occurs because of 
      feedback or because the available reservoir of 
      gas has been exhausted.
            %
      %
      We note that while the exact values of $V/\sigma$ and sSFR change from simulation to simulation, all simulations
      tend to predict qualitatively similar relations between 
      sSFR-$M_*$ \citep{Somerville2015} and $V/\sigma$ vs.
      $M_*$ \citep{vandeSande2019}.

   \subsection{Two distinct regimes for the IHSC and
               observational prospects in the local Universe.}
   \label{sec:regimes}
     Our results suggest that there are two broad, distinct,
     populations of halos, with a transition at 
     $\mstot \sim 10^{12}\,\msuni$.
     At $\mstot < 10^{12}\,\msuni$, properties of central 
     galaxies are highly correlated with $\fmihsc$, with more
     dispersion dominated galaxies being a strong indicator for
     high $\fmihsc$. 
     Weaker correlations are seen with star formation activity,
     likely rising from the more fundamental correlation with
     $V/\sigma$.
     Halo properties at these masses do not seem to be good
     indicators of the overall $\fmihsc$.
     At  $\mstot > 10^{12}\,\msuni$ a very different picture
     emerges. All centrals at these masses are similar: they
     have little star formation activity and are dispersion
     dominated, and so they offer little insight into the IHSC.
     Instead, the dynamical state of the correlations studied
     are readily available in the observations and hence offer
     a great opportunity to test our predictions.
     Moreover, observations have so far sampled only a handful
     of galaxies, which is insufficient to unveil a correlation
     like the one reported here.
     Surveys such as those possible with Hyper-Suprime Cam
     \citep{Miyazaki2012} and LSST \citep{Robertson2017} are likely
     to change this, because their deep photometry over half the sky will
     provide the required statistics and data quality. 

  \section{Temporal evolution of the IHSC}
  \label{sec:ihscevol}
    So far we have studied the mass content in the IHSC and how it 
    correlates with observable properties of its system at $z = 0$.
    We have, however, not yet explored the temporal information
    available from the simulation.
    In this section, we explore the evolution of the $\fmihsc$
    of the entire Horizon-AGN population with the aim of understanding
    how different regions of the $\mfsm$ plane are populated 
    as a function of time.
    Additionally, we study the evolution of systems of interest,
    specifically galaxy clusters and Milky Way-mass systems,
    using their individual evolutionary paths.

  \subsection{Individual paths}
  \label{sec:paths}

    In this section we follow the evolution of individual 
    systems through cosmic time by tracking their location in 
    the $\mfsm$ plane.
    We focus on three sets of objects of interest: Milky Way-mass
    systems with low and high $\fmihsc$ and galaxy clusters.

      \begin{figure*}
        \includegraphics[width=\textwidth]{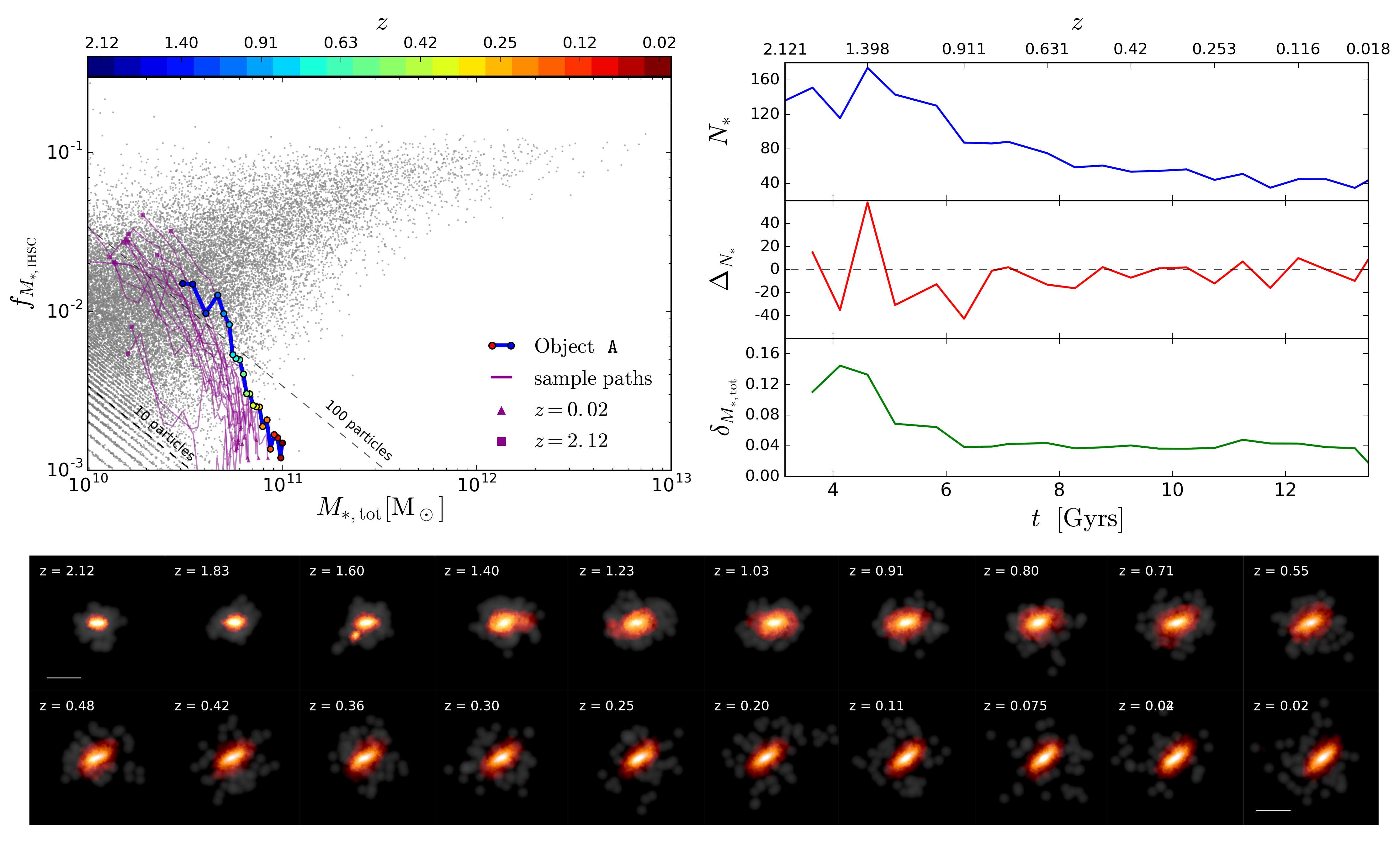}
        \caption{Evolution of Milky Way-mass systems with low $\fmihsc$.
                 \emph{Top left panel.} Evolutionary paths of a sample of galaxies
                 (thin magenta lines) from $z = 2.12$ (square symbols) to
                 $z = 0$ (triangle symbols).
                 In blue is highlighted the path of object \texttt{A}, with 
                 its location at each snapshot denoted by symbols coloured according
                 to redshift.
                 In the background the $z=0$ $\mfsm$ relation is shown as
                 scatter points; diagonal dashed lines delimit regions where the
                 IHSC is composed by 10 and 100 particles, as labelled.
                 \emph{Top right panels.} Number of particles in the IHSC,
                 $N_{*,\mathrm{IHSC}}$, difference of $N_{*,\mathrm{IHSC}}$ 
                 between snapshots,$\Delta_{N_*}$, and relative difference 
                 of $\mstot$, 
                 $\delta_{\mstot} = \mstot(t+\Delta t)/\mstot(t) - 1$, as
                 a function of time.
                 \emph{Bottom panels.} Stellar surface density projections
                 showing the evolution of the IHSC (gray) and the stellar mass
                 in galaxies (red). Brighter colours denote higher densities.
                 Colour intensities are different for each component to highlight
                 their distribution.
                 Systems with $\mstot \simeq 10^{11}\,\msuni$ and 
                 $\fmihsc \simeq 10^{-3}$ are characterised by having a rather
                 quiescent accretion history.
                 The absence of mergers reduces the mass deposited into the IHSC, 
                 that in addition to the mass growth of the galaxy through the
                 formation of new stars decreases the $\fmihsc$ of these systems.
                 See text for detailed discussion.
                 }
        \label{fig:ihsc_evol_path_mw_lf}
      \end{figure*}

    We follow the evolution of each system by tracking its 
    central galaxy across snapshots using 
    {\sc treefrog} \citep{Poulton2018,Elahi2019b}, a code that
    constructs merger trees for simulations.
    The progenitor of each galaxy is defined as the structure
    that shares most of its particles in the previous snapshot.
    This is found by comparing IDs of the stellar
    particles, and computing a merit function 
    
    \begin{equation}
      \mathcal{M}_{ij} = \frac{N^2_\mathrm{sh}}{N_i N_j},
    \end{equation}
    
    \noindent where $N_i$ and $N_j$ are the total number of 
    particles in galaxies $i$ and $j$ respectively, and 
    $N_\mathrm{sh}$ is the number of particles that exist
    in both structures.
    The main progenitor is chosen to be the one with the
    highest $\mathcal{M}_{ij}$.
    Systems are traced by following the main progenitors
    of the system's most massive galaxy at $z = 0$, i.e.
    the one with the highest merit.
    %
    
      To understand the evolutionary paths, we first note
      that horizontal displacements in the $\mfsm$ plane tend to be towards higher $\mstot$ increments, 
      which can either be caused by the accretion of systems or
      by the creation of new stars from gas accreted on an already
      existing Interstellar Medium (ISM);
      negative horizontal displacements are infrequent, and
      are caused by other galaxies or systems `flying by'.
      Vertical positive displacements correspond to mass being
      deposited from galaxies in the system (either central or
      satellites) into the IHSC.
      Negative vertical displacements result from 
      either increments in $\mstot$ for a fixed $\mihsc$, or
      the `re-capture' of particles from the IHSC to galaxies
      at fixed $\mstot$.

    \subsubsection{Milky Way-mass - Low $\fmihsc$}
      In Fig.~\ref{fig:ihsc_evol_path_mw_lf} we show the
      evolutionary paths in the $\mfsm$ plane for a sample
      of systems with $\mstot \simeq 10^{11}\msuni$ and 
      $\fmihsc \simeq 10^{-3}$ at $z=0$.
      At $z=2.12$ (magenta squares) these systems have 
      total stellar masses between $\mstot \simeq 10^{10}\,\msuni$
      and $10^{10.5}\,\msuni$, and a $\fmihsc \simeq 0.03$.
      Their evolution is mainly characterised by displacements
      towards higher $\mstot$ and lower $\fmihsc$.
      We highlight the evolution of object \texttt{A} (see
      Section~\ref{sec:ihscdef}) shown as a thick blue line, coloured
      symbols indicate its location in the plane at different
      redshifts, as labelled; 
      in the bottom panels we show the projected stellar density of
      the IHSC in grey, and the stellar mass in galaxies particles
      in red\footnote{Note that the colour intensity
      differs for the IHSC and for galaxies. This is done 
      to better highlight the IHSC, which tends to have a low number
      of particles. For comparison, the visualization of 
      Fig.~\ref{fig:ihscvis} has the same colour and intensity palette
      for both components.}.

      \begin{figure*}
        \includegraphics[width=\textwidth]{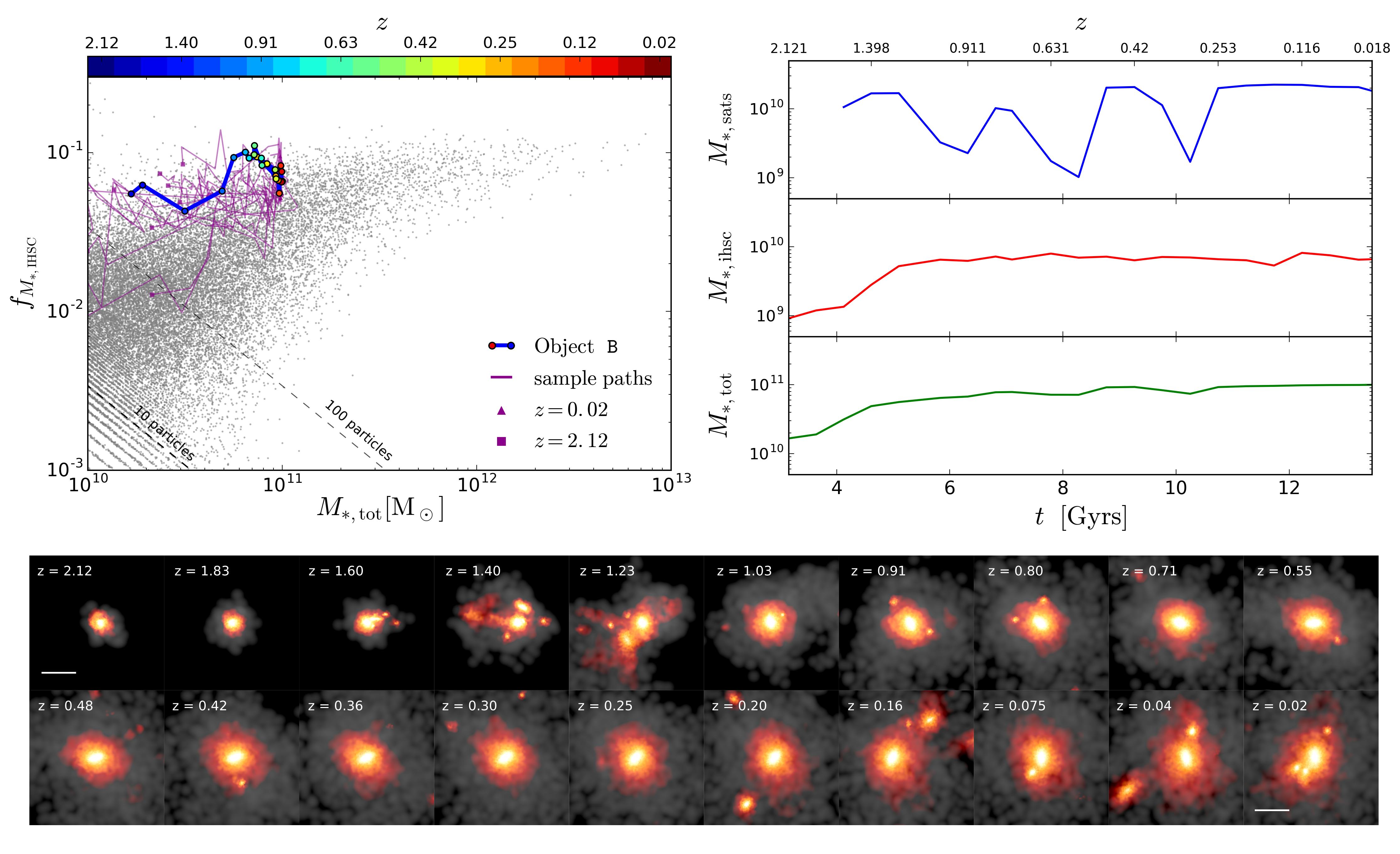}
        \caption{Same as Fig.~\ref{fig:ihsc_evol_path_mw_lf} but for
                 Milky Way-mass galaxies with high $\fmihsc$.
                 Top right panels show the total stellar mass, $\mstot$,
                 IHSC mass, $\mihsc$, and the mass in satellites $\msats$,
                 as a function of time.
                 Visualizations in the bottom panel have the same spatial 
                 scale, colour palettes and intensities as the example
                 shown in Fig.~\ref{fig:ihsc_evol_path_mw_lf}.
                 Systems with high $\fmihsc$ have an active accretion history
                 which is responsible for depositing a large amount of 
                 mass into the IHSC.
                 The diversity of paths in the $\mfsm$ plane described by 
                 these systems is the result of the variety of infall epochs
                 and satellite masses that each system experiences.
                 }
        \label{fig:ihsc_evol_path_mw_hf}
      \end{figure*}

      This system starts with $\fmihsc \simeq 0.01$, decreasing
      for a couple of snapshots before showing a sudden growth, followed
      by a monotonic decrease reaching $\fmihsc \sim 0.001$.
      This behaviour is a consequence of the
      interaction between an infalling satellite with the central
      galaxy.
      At $z=1.6$ the dip in the track is caused by the accretion
      of the satellite which increases $\mstot$, as seen in 
      $\delta_{\mstot} = \mstot(t+\Delta t)/\mstot(t) - 1$
      (top right panels), without yet depositing material onto
      the IHSC.
      At a subsequent snapshot, the number of particles in the
      IHSC increases, as seen in $N_{*,\mathrm{IHSC}}$ and 
      $\Delta_{N_{*,\mathrm{IHSC}}}$, due to the disruption of
      stellar material from the interaction, causing the bump
      in the track.
      At subsequent times, $\fmihsc$ keeps decreasing up 
      until $z=0.02$, which is due to the decreasing 
      number of particles in the IHSC, and an ongoing increment
      of $\mstot$, i.e. positive values of $\delta_{\mstot}$, 
      due to continuing star formation in the galaxy.
      The decreasing number of particles in the IHSC is also
      explained by mass being `re-accreted' onto the galaxy
      from the IHSC as the system relaxes due to the particular
      orbit that the infalling satellite had.
      This is in agreement with \citet{Penarrubia2006}
      and more recently \citet{Karademir2019}, who have shown
      that stellar debris from the disruption of galaxies on
      orbits close to the plane of the disk are expected to
      relax into an extended rotating disk;
      this extended disk would therefore occupy a similar region
      in phase-space and therefore be assigned to the galaxy
      in our method.
      Finally, we note that the bumps displayed by all systems
      at later times are caused by the low number of particles
      that compose the IHSC, as is shown by $\Delta_{N_{*,\mathrm{IHSC}}}$.
      All of the other highlighted galaxies in magenta show 
      a similar behaviour, interacting with
      a single satellite and continuous star
      formation activity in the central.

    \subsubsection{Milky Way-mass - High $\fmihsc$}
    In Fig.~\ref{fig:ihsc_evol_path_mw_hf} we show evolutionary
    paths of systems with $\mstot \simeq 10^{11}\,\msuni$ and
    high $\fmihsc$ at $z = 0$.
    The paths that these galaxies display in the $\mfsm$ plane
    are characterised by large displacements and variations
    in the $\fmihsc$, a consequence of the interactions that
    these systems experience.
    Contrary to the low $\fmihsc$ cases, these systems show
    a large variety of paths across cosmic time, due to the
    diversity of accretion histories.
    This is seen in the visualization of the highlighted path
    (bottom panels), as well as the evolution of the mass in 
    satellites, $\msats$ (top
    right panels), where bumps across time represent the infall
    (peaks) and disruption (valleys) of satellites.
    These multiple interactions experienced by the central galaxy
    at different times result in a higher mass fraction in the 
    IHSC.
    For comparison, the spatial scale and colour
    schemes of the projections is the same as for the
    low $\fmihsc$ case in Fig.~\ref{fig:ihsc_evol_path_mw_lf}.
    These results are consistent with the relation observed
    between the $\fmihsc$ of a system and $V/\sigma$ of its
    central galaxy.
    Moreover, these results demonstrate the robustness of our method, 
    and highlight the advantage of defining the IHSC as the
    kinematically hot component in each system.

      \begin{figure*}
        \includegraphics[width=\textwidth]{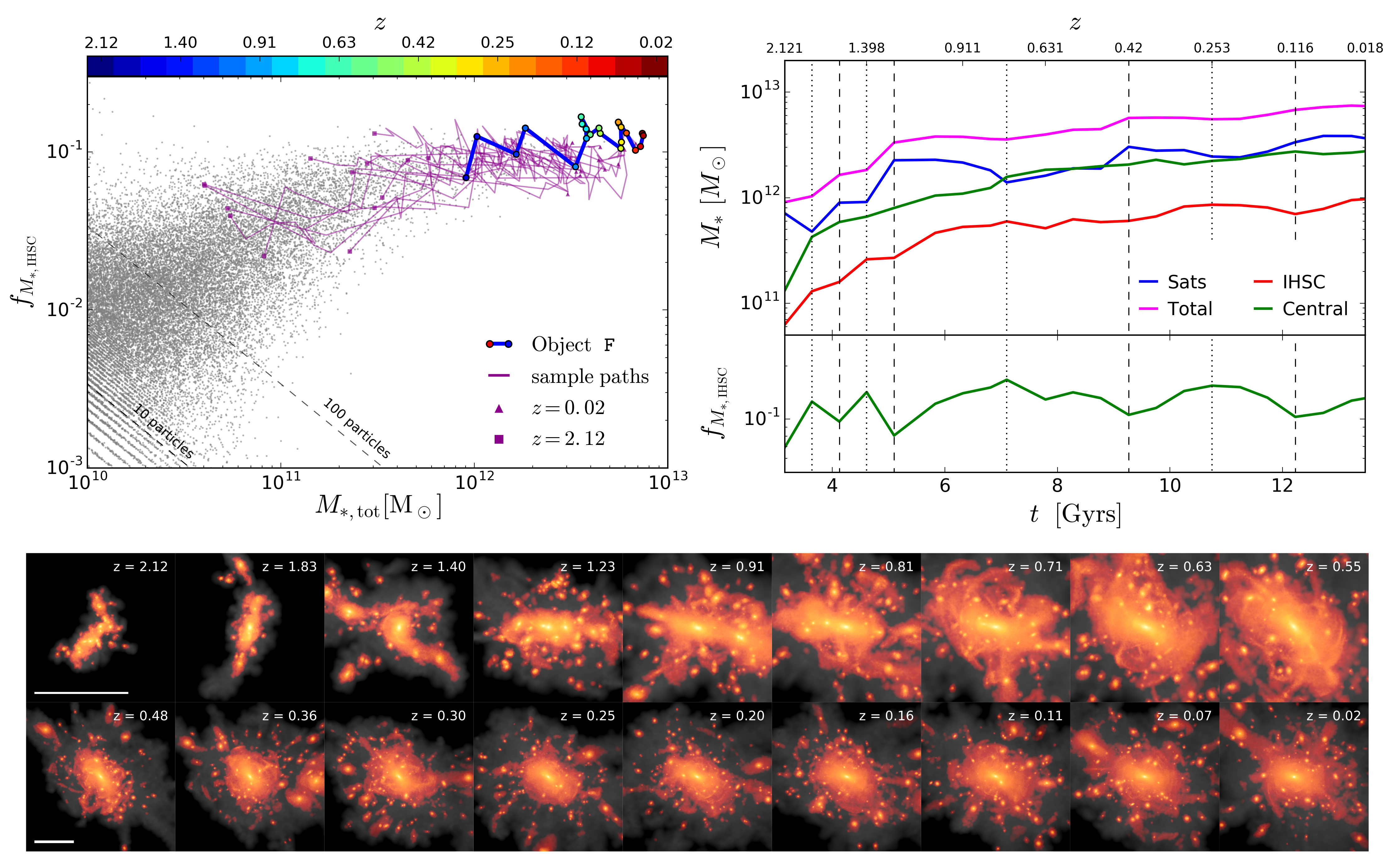}
        \caption{Same as Fig.~\ref{fig:ihsc_evol_path_mw_lf} but showing
                 the evolution of the IHSC at galaxy groups and clusters 
                 scales.
                 Note that the spatial scale is different for the top and
                 bottom row of the projections, in both cases the white
                 horizontal line has a length of 430 kpc.
                 Colour intensity is different from that of 
                 Figs.~\ref{fig:ihsc_evol_path_mw_lf} 
                 and~\ref{fig:ihsc_evol_path_mw_hf} for clarity 
                 purposes.
                 Top right panels show the evolution of the stellar mass 
                 in satellites, IHSC, central and total, as well as 
                 the evolution of the IHSC.
                 For reference vertical dotted (dashed) lines are shown
                 at peaks (valleys) of the $\fmihsc$.
                 The horizontal displacements in the $\mfsm$ plane
                 are the consequence of the interplay between the
                 disruption galaxies and the accretion of other groups
                 and clusters.
                 }
        \label{fig:ihsc_evol_path_clusters}
      \end{figure*}
    
    \subsubsection{Galaxy Clusters}
      In the top panel of Fig.~\ref{fig:ihsc_evol_path_clusters}, we 
      show evolutionary paths in the $\mfsm$ plane of systems
      with $\mstot > 10^{12}\,\msuni$ at $z = 0$, all of which have
      $\fmihsc > 5\times10^{-2}$ (magenta triangles).
      These objects display a variety of locations in the
      $\mfsm$ plane at $z=2.12$ (magenta squares), differing by
      $\sim 1$ dex in both $\mstot$ and $\fmihsc$ for this particular
      sample.
      Regardless of their initial location, all systems move towards
      higher $\mstot$ with time and display a series of peaks and valleys
      in the $\mfsm$ plane throughout their evolution, a consequence of
      the multiple
      interactions that drive the growth of these systems.
      For systems that at $z = 2.12$ have a high $\fmihsc$, the
      displacements are on average horizontal through cosmic time, as
      shown by the highlighted case.
      Systems that start with low $\fmihsc$ and $\mstot$
      show a tendency to move towards increasing $\fmihsc$ and 
      $\mstot$, until reaching $\mstot \simeq 10^{12}\,\msuni$ 
      where the displacements become on average horizontal.
      %
      
      While for Milky Way-mass systems this evolution can be
      described by the accretion of satellites onto a
      dominant central galaxy, for groups and
      clusters this evolution is more complex, because they grow mainly by
      accretion of entire smaller groups and clusters.
      These infalling systems carry with them their own IHSC,
      which after accretion is added to the IHSC of the larger
      system.
      During these events the total $\fmihsc$ is not expected
      to increase, simply because of how mass fractions add;
      for example, a cluster with an initial IHSC mass fraction
      of 10\% that accretes a group whose $\fmihsc$ is
      10\%, will have, post-accretion, the same mass fraction
      in the IHSC.
      Displacements towards higher $\fmihsc$ can therefore 
      only arise because mass is disrupted from galaxies and
      transferred to the IHSC.
      On average, systems describe horizontal paths because of
      the interplay between the disruption timescales of the galaxies
      already in the cluster and the accretion timescale of other
      clusters.
      %

      Overall these results indicate that, for systems with
      $\mstot > 10^{12}\,\msuni$, there is no apparent correlation
      between $\fmihsc$ and $\mstot$, and there should be no strong
      evolution of $\fmihsc$ with time.
      The lack of evolution of $\fmihsc$ with time is
    consistent with observational results of 
      \citet{Krick2007}, who found no trend between ICL fraction and 
      cluster mass for a sample of Abell clusters, and a weak or
      nonexistent evolution across redshifts. This has also been noted
      more recently by \citet{Montes2018}, using data from the
      Hubble Frontier Fields.
      In previous theoretical studies, \citet{Cui2014} found no
      correlation between cluster mass and the mass fraction
      in the ICL, using simulations that did not include AGN feedback
      and only when the ICL was defined using surface density 
      thresholds.
      This was not observed however when AGN feedback was included, 
      as well as when a dynamical definition of the ICL was used.
      This contradicts our results, and warrants further research
      at cluster masses.
      In a future paper we focus specifically on this regime using
      the Cluster-EAGLE \citep{Bahe2017,Barnes2017} suite
      (Ca\~nas et al. in prep.).

      \begin{figure*}
        \includegraphics[width=\textwidth]{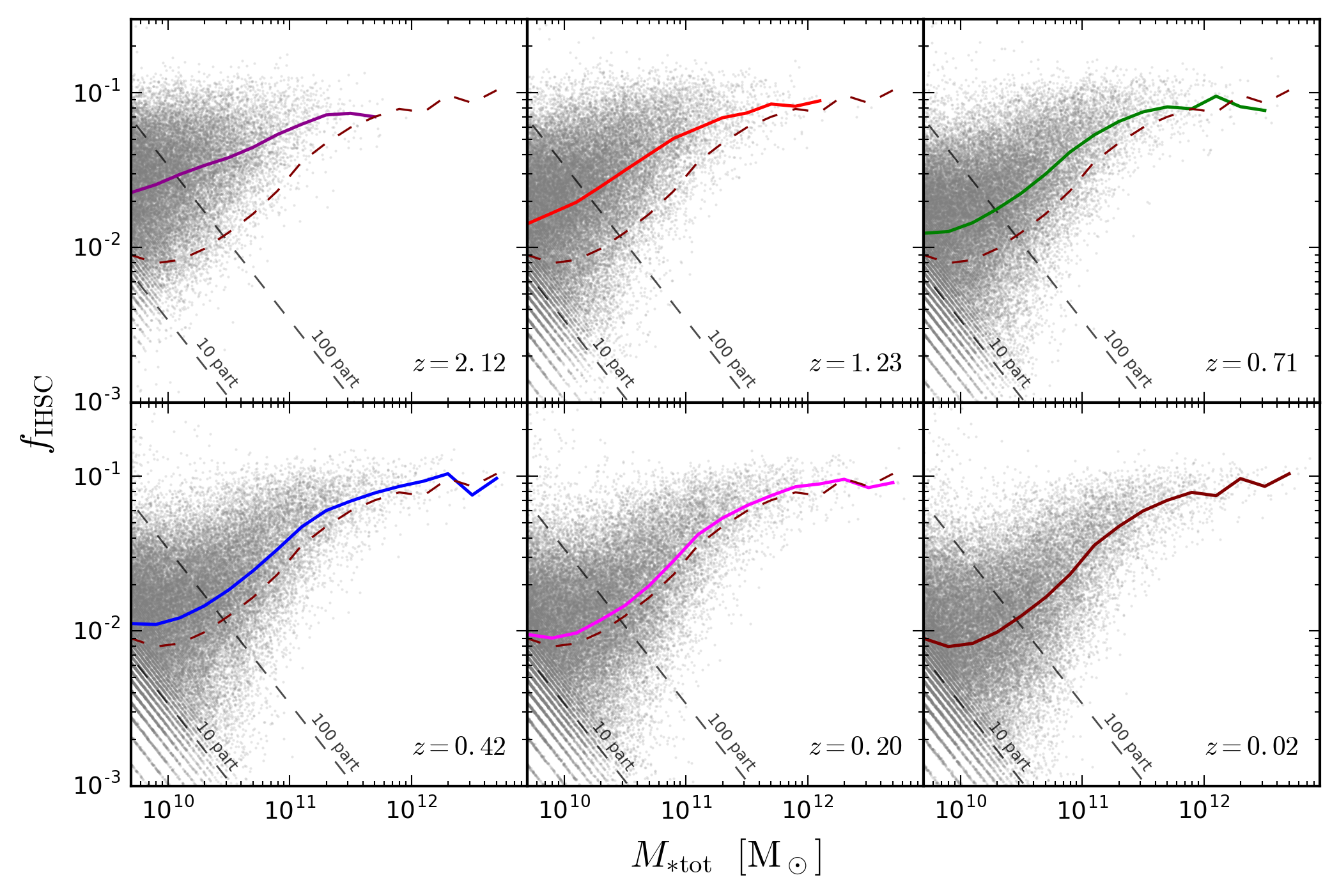}
        \caption{Temporal evolution of the $\mfsm$ relation from $z = 2.12$
                 to $z = 0.02$, as labelled.
                 Gray points are the measurements for individual
                 systems;
                 solid lines indicate the median $\fmihsc$ per mass bin,
                 and dashed line indicates the median at $z = 0$ for
                 reference.
                 Diagonal dashed lines indicate $\fmihsc$ limits for
                 IHSC composed 10 and 100 particles at a given
                 $\mstot$, as labelled.
                 } 
        \label{fig:ihsc_evol_pop}
      \end{figure*}

    \subsection{Galaxy population}
      In Fig.~\ref{fig:ihsc_evol_pop} we show the temporal
      evolution of the $\mfsm$ relation from $z = 2.12$ to
      $z = 0.02$.
      At $z=2.12$ the relation shows a mass dependence
      of the $\fmihsc$ with $\mstot$ up to 
      $\mstot \sim 2\times 10^{11}\,\msuni$,  where the
      relation appears to flatten.
      At $\mstot \simeq 10^{10}\,\msuni$ the scatter in the
      $\fmihsc$ can be greater than 1 dex, and decreases with
      increasing $\mstot$.
      At subsequent redshifts the evolution is characterised by
      a shift of the relation  towards higher $\mstot$, and is 
      accompanied by a steepening of the mean for 
      $\mstot < 10^{11.5}\,\msuni$, as well as a stronger
      distinction between a mass dependence of the $\fmihsc$ and 
      a weaker or non-existent one at the high mass end.
      This steepening of the relation is partially caused by
      systems populating low $\fmihsc$ regions at ever high
      $\mstot$, which, as seen in Section~\ref{sec:paths},
      correspond to systems that had a quiescent accretion
      history and whose IHSC ceased to grow.
      The other factor that causes this steepening is the
      horizontal displacement of systems with initially high
      $\fmihsc$, which also contribute to the flattening
      of the relation at $\mstot > 10^{11.5}\,\msuni$

  \section{Summary and conclusions}
  \label{sec:conclusions}
  
    We presented the first results of a new method to 
    define the Intra-Halo Stellar Component (IHSC) in
    cosmological hydrodynamical simulations.
    This method relies on robust identification of
    galaxies using an adaptive phase space structure
    finder {\sc velociraptor} \citep[][]{Canas2019,Elahi2019a}.
    The IHSC is then defined to be all the stellar
    material that is sufficiently distinct from galaxies
    in phase-space to be considered separate from them, i.e. a diffuse,
    kinematically hot stellar component.
    A critical feature of this method is the use of 
    \emph{local} properties of each system to separate
    the IHSC from the rest of the galaxies.
    This allows us to robustly define the IHSC 
    from Milky Way mass systems up to galaxy groups
    and clusters, and consequently to consistently follow
    the assembly of the IHSC through cosmic time.
    %

    A novelty of this method is that it is capable
    of producing consistent relations between the
    stellar mass of the central galaxy, $\mctrl$; the IHSC,
    $\mihsc$; and satellites, $\msats$, 
    independently of the phase-space density
    thresholds used, determined by the search parameter
    $f_{l_{x,\mathrm{6D}}}$.
    Modifying this parameter does not affect the
    shape of the relations, but only changes their
    zero-point.
    We argue that this feature allows the user to
    choose the $f_{l_{x,\mathrm{6D}}}$ value that 
    better matches observations of given characteristics, 
    e.g. surface-brightness limit, to make appropriate
    theoretical predictions.
    %

    We find that the $\fmihsc$ increases with the system's
    mass, on average, with the scatter being a strong
    function of the latter: low mass systems display variations
    of up to 2 orders of magnitude in their $\fmihsc$, 
    while the scatter gets systematically smaller, being
    0.3 dex at group masses.
    %

    We explored the nature of the scatter observed in 
    the $\mfsm$ relation at Milky Way masses
    by looking for possible correlations between
    properties of systems that are easily accessible in
    observations and their $\fmihsc$:
    \begin{description}
      \item[\textbf{Number of satellites.}] At 
            $\mstot > 10^{11}\,\msuni$, systems with higher
            $\nsats$ are expected to have a higher
            $\fmihsc$ at fixed $\mstot$, but at lower
            masses there is no evident
            correlation between $\nsats$ with $\fmihsc$.

      \item[\textbf{Dynamical age.}] In the case of $\mscnd/\mctrl$,
            it is only at $\mstot > 10^{12}\,\msuni$ that
            this quantity is indicative of the $\fmihsc$ in
            a system.
            At a fixed $\mstot$, dynamically younger systems 
            with higher $\mscnd/\mctrl$ have lower $\fmihsc$
            than older, more dynamically relaxed ones.

      \item[\textbf{Kinematic morphology.}] We found that
            $V/\sigma$ is the
            parameter that most strongly (anti)correlates
            with $\fmihsc$ at fixed mass in systems with 
            $\mstot < 10^{12}\,\msuni$.
            At Milky Way masses, we see a clear 
            transition from low $\fmihsc$ in systems with
            more rotationally supported galaxies, i.e. higher
            $V/\sigma$, to high $\fmihsc$ for those with a
            central galaxy mainly supported by
            dispersion, low $V/\sigma$.
            This is consistent with the picture that dispersion-supported galaxies have a more
            active accretion history and therefore a larger
            amount of mass deposited into the IHSC; this
            is in comparison with disk galaxies, which have on average a
            more quiescent accretion history.

      \item[\textbf{Star formation rate.}] At
            fixed $\mstot$, the SFR of the central galaxy 
            can be indicative of the expected $\fmihsc$ of the
            system, with lower $\fmihsc$ in systems with more
            star forming central galaxies.
            However, there is not a continuous transition from
            low $\fmihsc$ and high SFRs to high $\fmihsc$ and
            low SFRs, as strong as we obtained for $V/\sigma$ 
    \end{description}

    Our method also allows us to follow the evolution and
    assembly of the IHSC through cosmic time for individual
    objects as well as for the entire galaxy population.
    We explored individual paths in the $\mfsm$ plane for
    cases of interest:
    \begin{description}
      \item[\textbf{Milky Way mass - low $\fmihsc$.}] The evolution of
            these systems is characterised by smooth 
            displacements towards higher $\mstot$ and lower
            $\fmihsc$ which are the result of the absence of
            accretion of satellites,  the continuous formation
            of new stars from gas, and the relaxation of
            IHSC stars into the outskirts of the growing
            central galaxy.
              
      \item[\textbf{Milky Way mass - high $\fmihsc$}] These galaxies 
            display a variety of evolutionary paths in the
            $\mfsm$ plane.
            Contrary to low $\fmihsc$ cases, these systems 
            have different locations in the plane at $z=2$, 
            but finish with similar $\mstot$ and $\fmihsc$
            at $z = 0$.
            Their paths are characterised by increments in the
            $\fmihsc$, which are a consequence of 
            episodes of satellite accretion and interactions,
            that occur at different epochs for each system.

      \item[\textbf{Galaxy groups and clusters}] The evolutionary
            paths of high-mass systems are characterised by a 
            series of increments and decrements in the $\fmihsc$.
            While for Milky Way mass systems the evolution 
            of these features are attributed to the accretion and
            disruption of individual satellites, for groups
            and clusters such features are caused by the accretion
            of smaller groups/clusters and the disruption of 
            all the members within those structures.
            Overall these systems display increasing $\fmihsc$
            with time until they reach a 
            $\mstot \sim 10^{11.5}\,\msuni$ and 
            $\fmihsc \simeq 10-20$\%, from where the
            evolution indicates to be weak or nonexistent due
            to the interplay between group accretion episodes, 
            which decreases $\fmihsc$, and disruptions of
            individual galaxies that increase it.
    \end{description}

    The evolution of the galaxy population is  characterised
    by a steepening of the median $\fmihsc$ at 
    $\mstot < 10^{11}\,\msuni$, which is caused by galaxies with
    $\fmihsc$ at high redshifts that subsequently have a quiescent
    accretion history, populating the low $\fmihsc$ region in 
    increasing numbers as time goes by.
    For $\mstot > 10^{11}\,\msuni$, the population evolution
    shows a mild to neglible increment in the $\fmihsc$.
    For our preferred 6D finder parameters, the $\fmihsc$
    appears to peak around $\sim10$\%, however such value
    changes depend on the phase space density cut used.
    %

    Overall, the scatter displayed by the $\mfsm$
    relation is driven by the large diversity in accretion
    histories that galaxies can experience.
    The difference in the scatter observed at Milky 
    Way-masses and at groups and cluster scales is mainly
    due to the various factors that play a role in the
    growth and evolution of a Milky Way-mass system,
    compared to that of clusters, which are dominated by
    hierarchical growth.
    While some properties of the system can be used
    as indicators of the stellar content in the IHSC, the
    specific location in the $\mfsm$ plane ultimately
    depends on the specific assembly history a system has,
    because specific locations in the
    $\mfsm$ plane can be reached by a variety of paths,
    explaining why there is a non-negligible scatter even
    when studying sub-samples of the population with 
    more similar accretion histories.
    %
    
    The method and results presented in this paper 
    give insight into open questions about the IHSC, and
    provide a consistent picture of the properties of 
    the IHSC across a wide range of masses and its
    evolution throughout cosmic time.
    There is, however, still work to be done, more 
    specifically exploring in detail Milky Way mass
    system using higher resolution simulations, as
    well simulations that can resolve multiple
    massive galaxy clusters.
    Finally, a detailed and careful comparison between
    observations and different theoretical models is
    crucial to help us understand better the physics of
    galaxy formation.
    %

     
  \section*{Acknowledgements}
    The authors thank Aaron Robotham, Bob Abraham, Aaron Ludlow,
    Mireia Montes, Rhea-Silvia Remus, Rhys Poulton and the theory 
    and computing group at ICRAR for helpful discussions. 
    RC is supported by the MERAC foundation postdoctoral
    grant awarded to CL and by the Consejo Nacional de Ciencia
    y Tecnolog\'ia CONACYT CVU 520137 Scholar 290609 Overseas
    Scholarship 438594.
    CL is funded by a Discovery Early Career Researcher Award 
    (DE150100618). 
    Parts of this research were conducted by the Australian Research
    Council  Centre  of  Excellence  for  All  Sky  Astrophysics  in  
    3  Dimensions (ASTRO 3D), through project number CE170100013.
    This research is part of Spin(e) 
    (ANR-13- BS05-0005, http://cosmicorigin.org).
    This work has made use of the Horizon Cluster hosted by Institut 
    d'Astrophysique de Paris. We thank Stephane Rouberol for running 
    smoothly this cluster for us.
    %



\bibliographystyle{mnras}
\bibliography{references}



\appendix

  \section{IHSC extent and spherical apertures}
  \label{appndx:fixedapertures}
  
\begin{figure}
      \centering
      \includegraphics[width=\columnwidth]{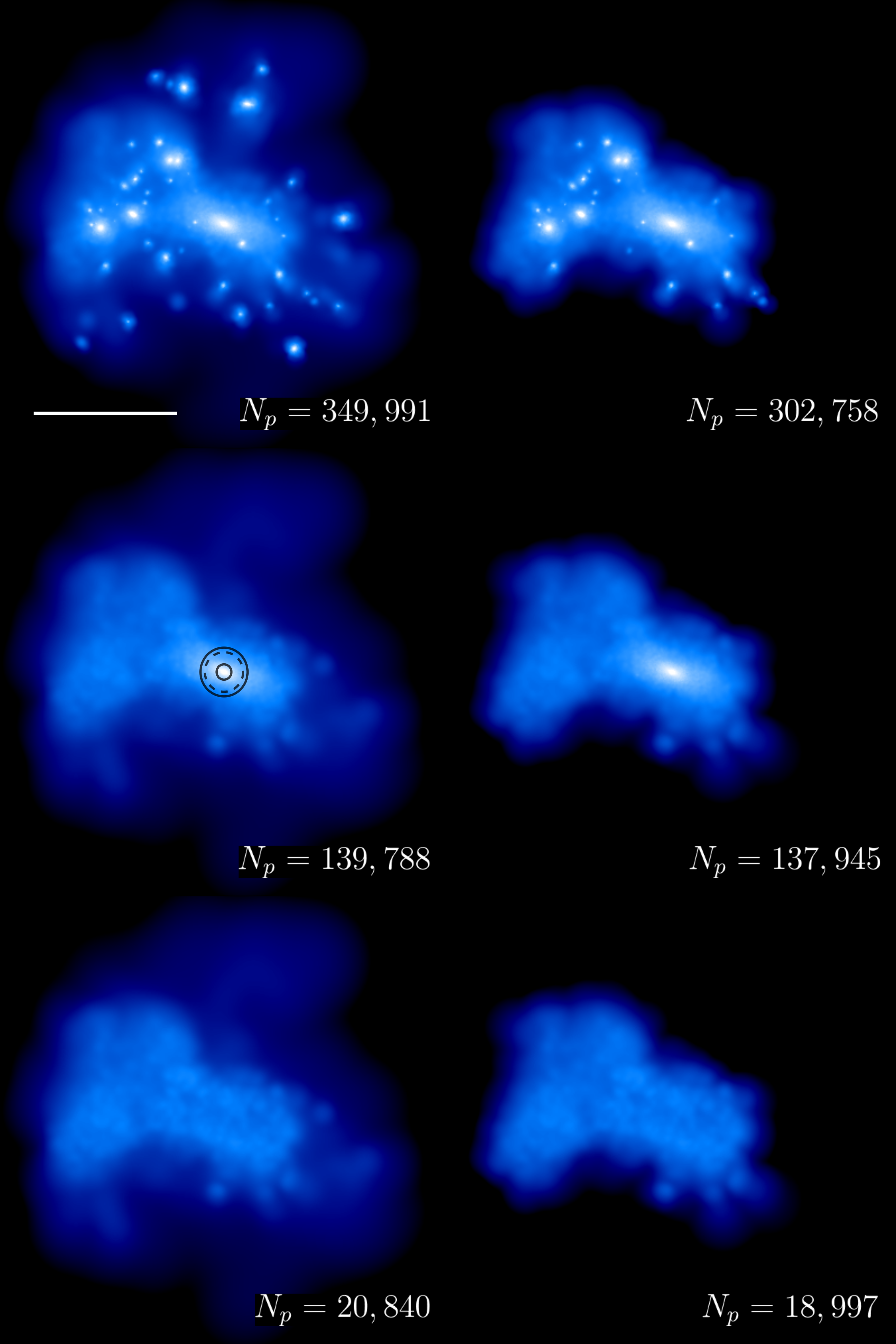}
      \caption{Surface density projections of total stellar content
               (top row), central galaxy and IHSC (CG+IHSC, middle
               row), and IHSC only (bottom row), for spherical
               overdensity $M_{200c}$ (left column) and 3DFOF
               conventions for a group in Horizon-AGN at $z=0$.
               The total number of stellar particles, $N_p$, in the
               object is indicated in each panel;
               solid white line represents a spatial scale of 
               570 kpc.
               Concentric circles denote the extent of spherical
               apertures used in the literature to separate the
               central galaxy from the IHSC.
               For the $M_{200c}$ CG+IHSC we also show spherical apertures
               of 30 and 100 kpc (solid circle), and $2\,R_{50}$ (dashed
               circle), for which the stellar material inside the
               aperture is considered as the central galaxy, and the outer
               to be the IHSC \citep[e.g.][]{Pillepich2018b,Elias2018}.
               See text for a detailed description.
              }
      \label{fig:spherical_apertures}
    \end{figure}

\begin{figure}
      \centering
      \includegraphics[width=\columnwidth]{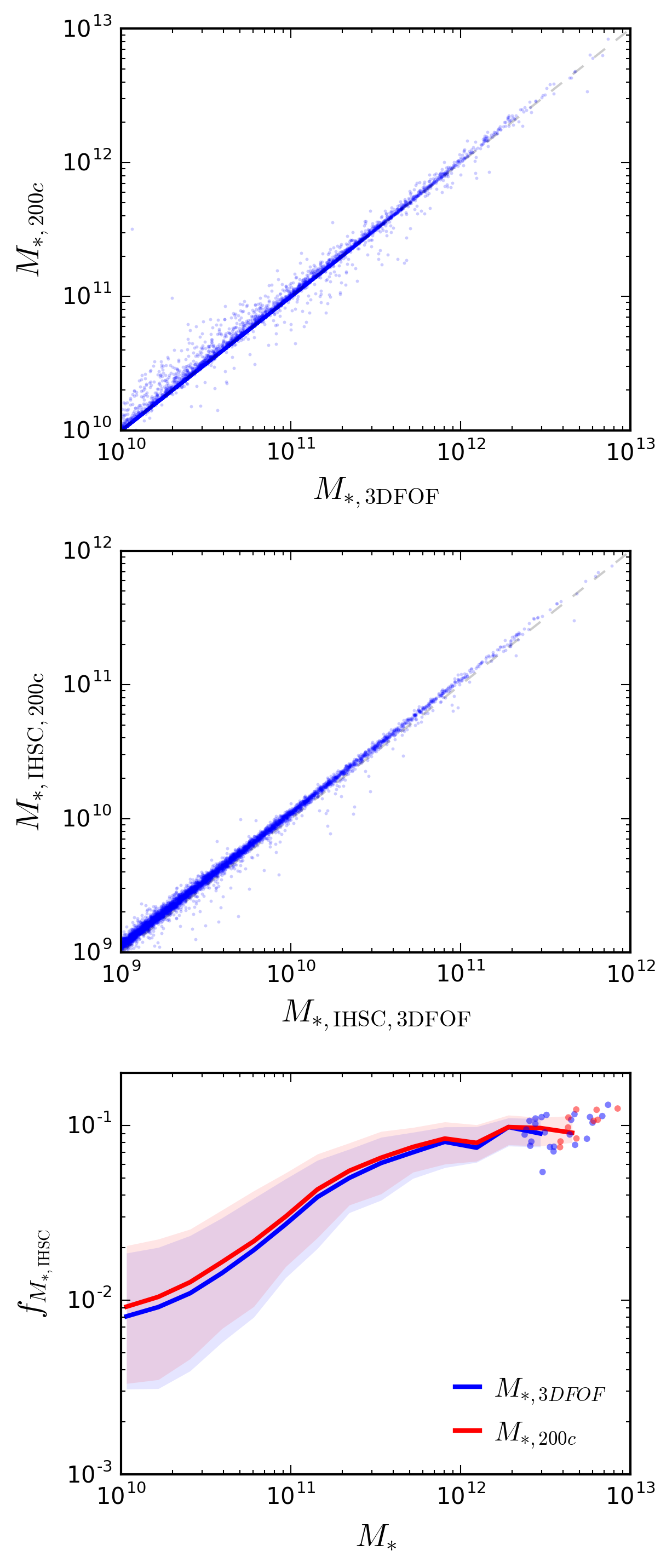}
      \caption{Total stellar mass content (top), mass in the 
               IHSC (middle), and $\fmihsc-M_*$ relation of
               3DFOF objects compared to its correspondent
               spherical overdensity of 200 times the critical
               density of the universe (bottom), for all 3DFOF
               objects in Horizon-AGN at $z=0$.
               Solid lines and shaded regions in the bottom panel
               represent the median, and extent of 16$^\mathrm{th}$
               and 84$^\mathrm{th}$ percentiles, respectively.
               The stellar mass content is in agreement for both
               conventions, displaying an almost 1-to-1
               correspondence and a consistent $\fmihsc-M_*$
               relation in the whole mass range.
               }
      \label{fig:ihsc_extent}
    \end{figure}

    Here we present how our definition of the IHSC compares
    to the one defined by spherical overdensity and spherical
    apertures widely used in the literature (see 
    Section~\ref{sec:intro} for a discussion).
    In Fig.~\ref{fig:spherical_apertures} we show surface 
    density projections of the stellar mass content using
    $M_{200c}$ (left column) and 3DFOF (right column) 
    mass conventions.
    Top row shows the total stellar mass, middle row
    the stellar mass in the central galaxy and IHSC (CG+IHSC),
    and bottom row the stellar mass in the IHSC only, as
    identified by {\sc velociraptor}.
    The total mass content (top row) shows that 3DFOF 
    convention does not capture the presence of some 
    satellites in the vicinity of the central galaxy, as
    well as some diffuse stellar mass, differing by
    $\sim 15$\%.
    This difference is similar to
    the excess of $\sim 10\%$ mass content in the $M_{200c}$
    IHSC and 3DFOF definitions.
    For this particular example, the mass fraction in the IHSC
    is $\fmihsc = 0.059$ for $M_{200c}$ and $\fmihsc = 0.062$.
    This behaviour is consistent for the entire population of
    objects resolved in the Horizon-AGN simulation, as
    can be seen in Fig.~\ref{fig:ihsc_extent}, where the total
    stellar mass content, $M_{*,\mathrm{tot}}$ (top panel), and
    the mass in the IHSC, $M_{*,\mathrm{IHSC}}$ (middle panel),
    follow an almost 1-to-1 correspondence with only notable
    variations at $M_* < 10^{11}\,\msuni$.
    Finally, the $\fmihsc-M_*$ relation (bottom panel) is 
    consistent in shape and amplitude at all stellar masses
    for both conventions.

    \begin{figure}
      \centering
      \includegraphics[width=0.93\columnwidth]{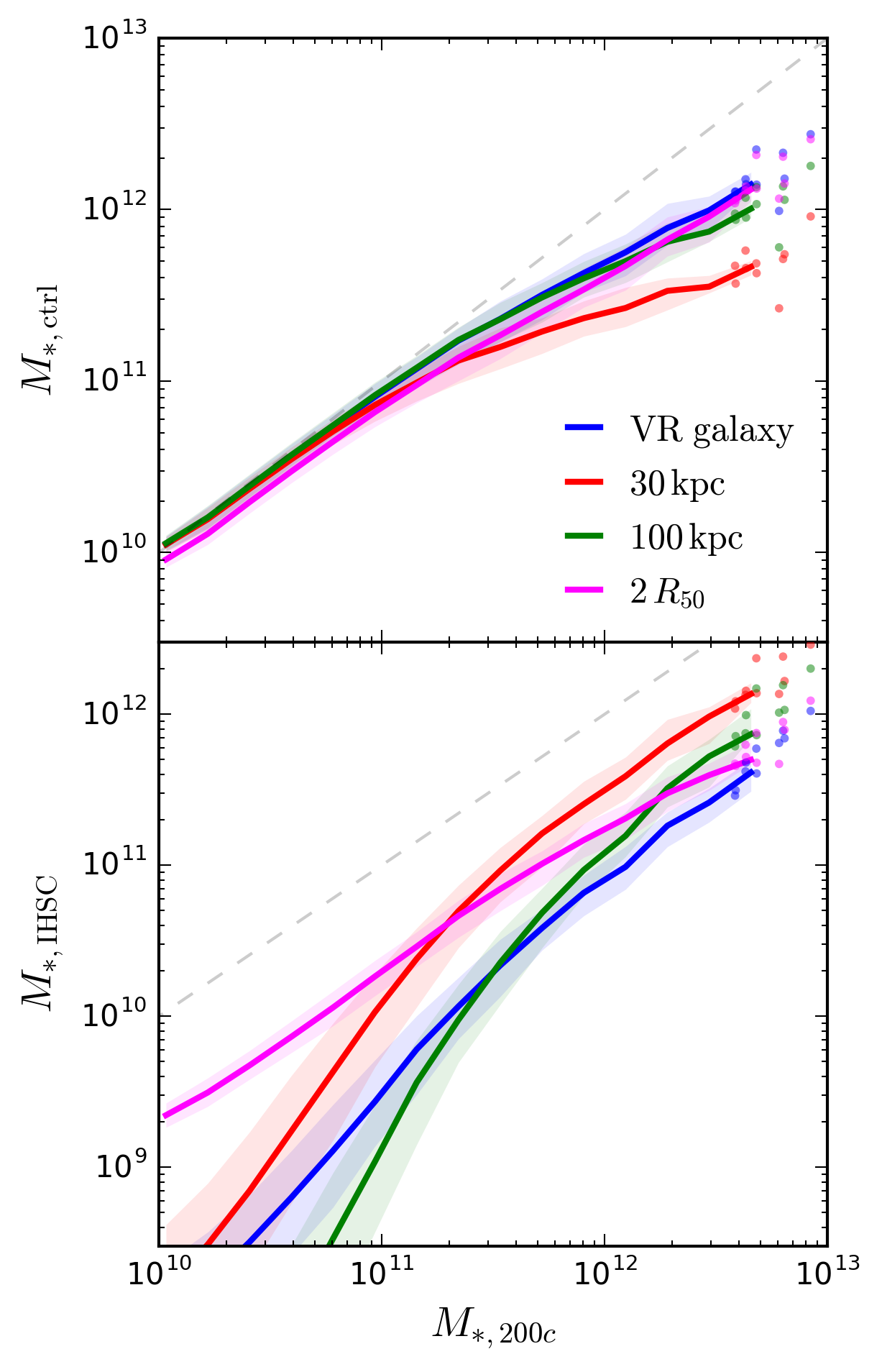}
      \caption{Stellar mass of the central galaxy (top) and IHSC 
               mass (bottom) as a function of total stellar mass
               within spherical a overdensity for all systems in
               Horizon-AGN at $z=0$.
               }
      \label{fig:m200c_mihsc_mctrl}
    \end{figure}

    \begin{figure}
      \centering
      \includegraphics[width=0.9\columnwidth]{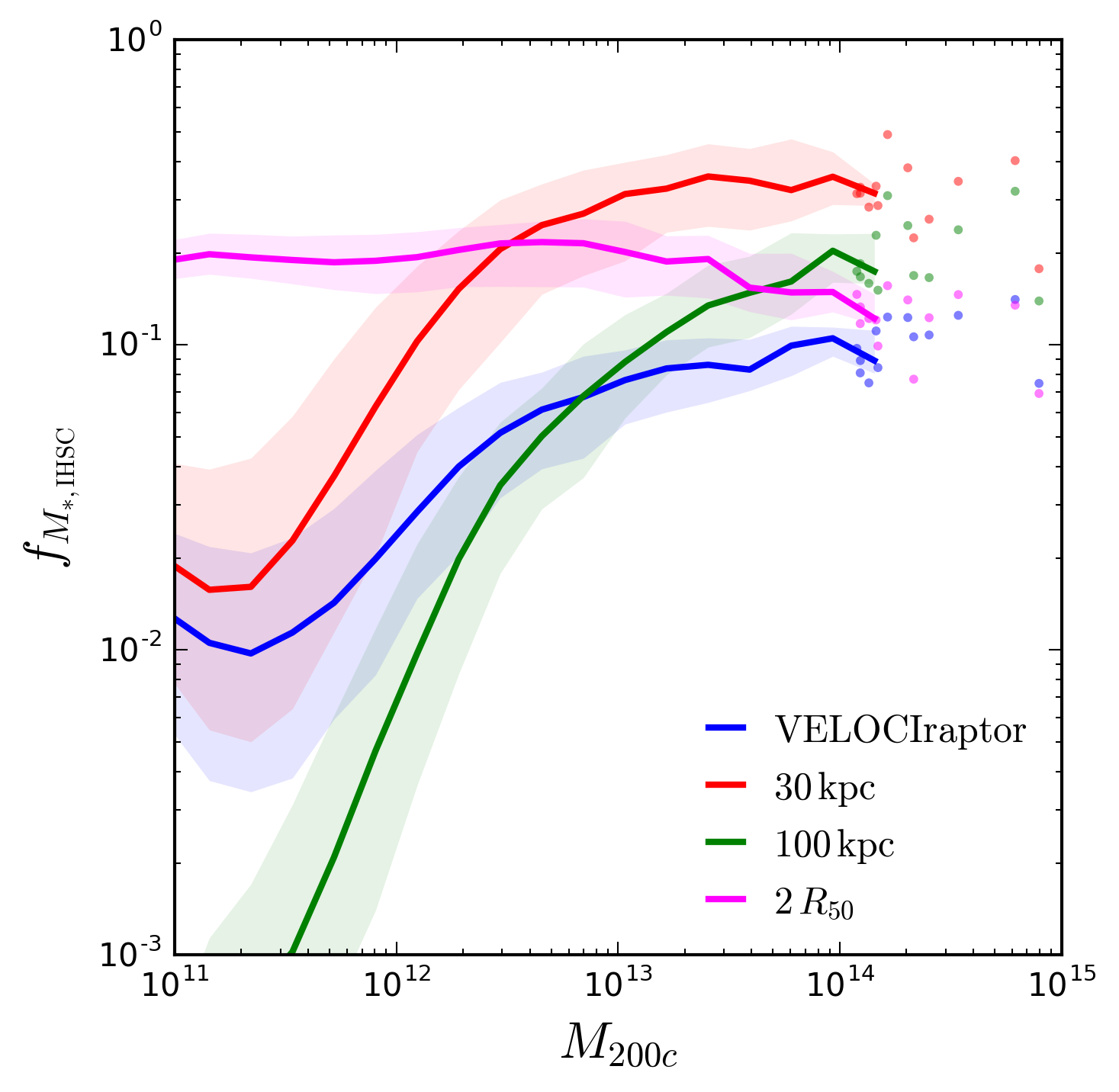}
      \caption{Mass fraction in the IHSC, $\fmihsc$, as a function of
               total halo mass $M_{200c}$ (bottom).
               }
      \label{fig:fmihsc_m200c}
    \end{figure}

    In the middle panel of Fig.~\ref{fig:spherical_apertures},
    we show the combination of CG+IHSC components, along
    with concentric circles representing spherical apertures
    of 30 and 100 kpc (solid circles) and $2\,R_{50}$ (dashed
    circle) used in the literature to separate the 
    central galaxy from the IHSC 
    \citep[e.g.][]{Pillepich2018b,Elias2018}.
    We show in the top panel of Fig.~\ref{fig:m200c_mihsc_mctrl} 
    the stellar mass of the central galaxy, $\mctrl$ as a function of 
    total stellar mass inside spherical over-density, $M_{*,200c}$;
    note that for the {\sc velociraptor} measurement we use the
    definition of the bottom left panel of 
    Fig.~\ref{fig:spherical_apertures}, instead of the 3DFOF
    for consistency.
    All definitions show a consistent relation between $\mctrl$
    and $\mvir$ for $\mvir < 10^{11}\,\msuni$, which is 
    expected as the mass budget is contained mainly in the central
    galaxy, which at this masses can be well defined by spherical
    apertures of 30 and 100 kpc.
    At larger masses the relations deviate from the 1-to-1 relation.
    This is in part because most of the mass comes from satellites,
    which is particularly noticeable for the 30 kpc aperture as the
    central galaxy is also likely to extend beyond this aperture.
    A similar trend is visible at $\mvir > 10^{12.5}\,\msuni$ for 
    the 100 kpc aperture for the same reason.
    The trend displayed by {\sc velociraptor} and $2\,R_{50}$ 
    appears to indicate that the {\sc velociraptor} relation flattens
    a little at higher masses, while the slope of $2\,R_{50}$ 
    is steeper, however, statistics are low to make a strong
    conclusion.
    Contrary to the $\mctrl$, the relation between $\mvir$ and 
    $\mihsc$ is different for all definitions at all mass ranges.
    The $2\,R_{50}$ relation has a power-law index $< 1$ at all masses.
    In the case of the 30 kpc aperture, there is an evident change
    in the power-law index which goes from being $\geq 1$ at 
    $\mvir < 2\times10^{11}\,\msuni$
    to $\approx 1$ for higher masses.
    The 100 kpc aperture also shows a change in the power-law index
    at a similar mass, but with a value $> 1$;
    finally for our method the initial slope is slightly $> 1$
    for $\mvir < 10^{11.5}\,\msuni$ and $\sim1$ at higher masses.
    %

    Lastly, in Fig.~\ref{fig:fmihsc_m200c} we show the $\fmihsc$
    as a function of the total halo mass $\mvir$.
    This plot is complementary to that of Fig.~\ref{fig:fmihsc_ms200c}.
    While the $M_{*,200c}$ relation is observable for low-mass systems,
    the $M_{200c}$ is useful for theoretical predictions, and can also
    be estimated via lensing at cluster scales.
    The behaviour of both relations is very similar, which is
    expected from the 1-to-1 comparison shown in 
    Fig.~\ref{fig:ihsc_extent}.
    %


\bsp	
\label{lastpage}
\end{document}